\documentclass[aps,showpacs,amsmath,amssymb,twocolumn,pra,superscriptaddress,notitlepage]{revtex4-2}

\usepackage{qcircuit}
\usepackage{amsmath,amssymb,amsthm,mathtools}
\usepackage{bm}
\usepackage{graphicx}
\usepackage{subfigure, epsfig}
\usepackage{braket}
\usepackage{enumerate}
\usepackage{color}
\usepackage{algorithm}
\usepackage{algorithmic}
\usepackage{comment}
\usepackage{appendix}
\usepackage{here}
\usepackage{tabularx}
\usepackage{physics}
\usepackage{mathrsfs,amsfonts,dsfont}

\newcommand{\pd}[3]{
 \if 1#1 \frac{\partial #2}{\partial #3}
 \else \frac{\partial^{#1} #2}{\partial #3^{#1}}\fi}
 \newcommand{\od}[3]{
 \if 1#1 \frac{{\mathrm d} #2}{{\mathrm d} #3}
 \else \frac{{\mathrm d}^{#1} #2}{{\mathrm d}#3^{#1}}\fi}

\begin{document}

% -------------------- TITLE --------------------

\title{Systematic construction of digital autonomous quantum error correction \\
for state preparation and error suppression via conditional Gaussian operations}

% ------------ AUTHORS AND AFFILIATIONS ----------

\author{Keitaro Anai}
\affiliation{Department of Applied Physics, School of Engineering, The University of Tokyo,
7-3-1 Hongo, Bunkyo-ku, Tokyo 113-8656, Japan}

\author{Suguru Endo}
\email{suguru.endou@ntt.com}
\affiliation{NTT Computer and Data Science Laboratories, NTT Inc., Musashino 180-8585, Japan}
\affiliation{NTT Research Center for Theoretical Quantum Information, NTT Inc. 3-1 Morinosato Wakanomiya, Atsugi, Kanagawa, 243-0198, Japan}

\author{Shuntaro Takeda}
\affiliation{Department of Applied Physics, School of Engineering, The University of Tokyo,
7-3-1 Hongo, Bunkyo-ku, Tokyo 113-8656, Japan}

\author{Tomohiro Shitara}
\email{tomohiro.shitara@ntt.com}
\affiliation{NTT Computer and Data Science Laboratories, NTT Inc., Musashino 180-8585, Japan}
\affiliation{NTT Research Center for Theoretical Quantum Information, NTT Inc. 3-1 Morinosato Wakanomiya, Atsugi, Kanagawa, 243-0198, Japan}

% -------------------- ABSTRACT --------------------
% 162/500 words
\begin{abstract}
In continuous-variable quantum computing, autonomous quantum error correction (QEC) can dissipatively steer a noisy quantum state into a target state or manifold, enabling robust quantum information processing without explicit syndrome measurements and feedback. Here, we propose a nullifier-based digital autonomous QEC enabled by conditional Gaussian operations. By designing jump operators for target nullifiers and compiling the resulting Lindbladian into a Trotterized sequence of elementary conditional Gaussian operations, we demonstrate two use cases: (i) deterministic preparation of non-Gaussian resource states for universal computation, including finitely squeezed cubic phase states and approximate trisqueezed states, and (ii) autonomous suppression of dephasing error for cat and squeezed cat states. We provide explicit gate decompositions for the required conditional Gaussian operations and numerically evaluate the performance under realistic imperfections, including photon loss in the bosonic mode and ancillary-qubit decoherence. Our results clarify the resource requirements and trade-offs—such as circuit depth, time-step choices, and the required set of conditional Gaussian operations—for scalable, gate-level implementations of autonomous state preparation and error suppression.

\end{abstract}

\maketitle

% -------------------------------------------------------
\section{Introduction}
Quantum computers are expected to provide significant speedups for practical tasks such as quantum simulation~\cite{georgescu2014quantum, bauer2020quantum}, machine learning~\cite{mitarai2018quantum,biamonte2017quantum}, and integer factorization~\cite{365700}. While the qubit-based quantum computing paradigm has been the primary focus in this field, continuous-variable (CV) quantum computing offers potential advantages because it leverages infinite-dimensional Hilbert spaces for computation~\cite{adesso2014continuous, cai2021bosonic, joshi2021quantum, ths:CV-QAOA, ths:CV-QML, fukui2022building}.

The remarkable advantage of CV is a hardware-efficient implementation of quantum error correction (QEC). Both rotation symmetry-based and translation symmetry-based QEC have seen a break-even point, i.e., the coherence time of error-corrected qubits exceeds that of uncorrected qubits~\cite{Ofek2016, Hu2019, Ni2023BeatingBreakEven, sivak2023real}. In matter systems, such as superconducting and ion trap systems, autonomous QEC has been primarily studied~\cite{PhysRevLett.125.260509}. This approach aims to dissipate the noisy quantum state into the target state or manifold by designing a dissipator, thereby eliminating the need for syndrome measurements and feedback operations.

Besides the progress in CV-QEC, significant efforts have been devoted to expanding the set of available operations in CV quantum computing. In this context, conditional squeezing has recently been proposed~\cite{ayyash2024driven, del2024controlled}. While few applications of conditional squeezing have been proposed~\cite{ayyash2024driven, q15h-nmq9}, its broader potential has yet to be fully explored.

\begin{figure}[t]
    \centering
    \includegraphics[width=1.0\linewidth]{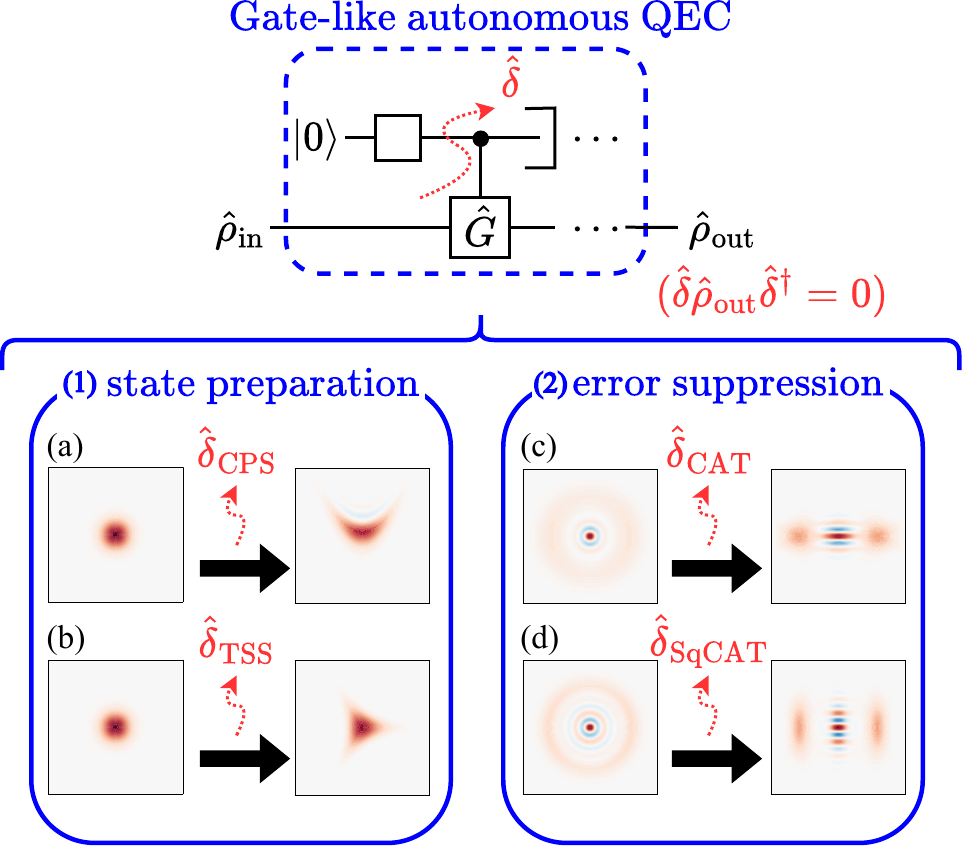}
    \caption{\textbf{Concept of autonomous QEC.} Given an operator $\hat{\delta}$, our goal is to produce an output state $\hat{\rho}_{\mathrm{out}}$ satisfying $\hat{\delta}\hat{\rho}_{\mathrm{out}}\hat{\delta}^\dagger=0$. The autonomous QEC mimics the Gorini–Kossakowski–Sudarshan–Lindblad (GKSL) dynamics with jump operator $\hat{\delta}$ so that, in the long-time limit, the system relaxes to $\hat{\rho}_{\mathrm{out}}$. With this autonomous QEC, we consider two use cases in this paper: (1) preparing a target quantum state and (2) suppressing an error of a target state. For the state preparation, we consider the preparation of (a) the cubic phase state with the dissipator $\hat{\delta}_{\mathrm{CPS}}$ and the preparation of (b) the trisqueezed state with the dissipator $\hat{\delta}_{\mathrm{TSS}}$. For the error suppression, we consider the suppression of the dephasing error of (c) the cat state with the dissipator $\hat{\delta}_{\mathrm{CAT}}$ and (d) the squeezed cat state with the dissipator $\hat{\delta}_{\mathrm{SqCAT}}$. $\hat{G}$ indicates the Gaussian operations, which are implemented easily in the experiment.}
    \label{fig:Concept}
\end{figure}

In this work, we propose a nullifier-based digital dissipative synthesis approach to autonomous QEC, and consider two use cases as shown in Fig.~\ref{fig:Concept}.
The first use case is the preparation of resource states for universal quantum computing, i.e., cubic phase states and approximate trisqueezed states. We design a digital-type quantum circuit for resource-state preparation via autonomous QEC.
We show that conditional Gaussian operations enable autonomous QEC that dissipatively steers an initial state into the cubic phase state~\cite{hillmann2020universal,eriksson2024universal} and the approximate trisqueezed state~\cite{zheng2021gaussian}.
Compared with prior platform-specific proposals that realize cubic phase states by engineering explicit higher-order nonlinearities~\cite{hillmann2020universal,eriksson2024universal}, our approach compiles the target dissipator digitally into a Trotterized sequence of conditional Gaussian operations. In this sense, a standalone cubic phase gate is not required as an elementary operation, and the resource cost is instead captured by the depth and structure of the resulting conditional Gaussian operations.
The same compilation idea also provides a unified dissipative route to approximate trisqueezed states.

The second use case is the error suppression of certain QEC codes. Concretely, we show that conditional Gaussian operations suppress the error of the cat codes~\cite{PhysRevA.59.2631, Mirrahimi_2014}. We also show that the error of the squeezed variant of cat codes, i.e., squeezed cat codes~\cite{PhysRevA.107.032423, PhysRevA.106.022431}, can be suppressed with a similar quantum circuit. While autonomous QEC for cat and squeezed cat codes has been extensively studied~\cite{Mirrahimi_2014, Gertler2021, PhysRevX.8.021005, PhysRevA.107.032423, shitara2025exploitingtranslationalsymmetryquantum}, previous approaches typically rely on the direct engineering of specific Hamiltonians. In contrast, our scheme decomposes the required protocol into elementary conditional Gaussian operations, such as displacement and squeezing operations, which we expect to enable a more digital and scalable implementation with greater flexibility in circuit design.

The rest of this paper is organized as follows. In Sec.~\ref{sec:cv_states}, we review CV quantum computing and, as a new contribution of this work, introduce the nullifiers corresponding to finitely squeezed cubic phase and trisqueezed states. In Sec.~\ref{sec:gate_level}, we review the autonomous QEC. In Sec.~\ref{sec:physicalimplementation}, we present concrete circuit constructions for specific states introduced in Sec.~\ref{sec:cv_states}. In Sec.~\ref{Sec:UseCasesAnd}, we provide two use cases, namely state preparation and error suppression. Section~\ref{sec:numerics} reports numerical simulations. Finally, Sec.~\ref{sec:Discussion} concludes the paper with a discussion.

% (Introductory text intentionally minimal; placeholder content retained per author instruction.)

% -------------------------------------------------------
\section{Continuous-variable quantum states}
\label{sec:cv_states}
Here, we introduce the CV quantum states considered in this work: the cubic phase state, trisqueezed state, cat state, and squeezed cat state. We consider a single harmonic oscillator with annihilation and creation operators $\hat{a}$ and $\hat{a}^\dagger$ obeying $[\hat{a},\hat{a}^\dagger]=1$. The quadrature operators are $\hat{x}=(\hat{a}+\hat{a}^\dagger)/\sqrt{2}$ and $\hat{p}=-i(\hat{a}-\hat{a}^\dagger)/\sqrt{2}$.

We refer to an operator $\hat{\delta}$ as a \emph{nullifier} of a quantum state $\ket{\psi}$ if $\hat{\delta}\ket{\psi}=0$. Under a unitary operation $\hat{U}$, the state transforms as $\ket{\psi}\mapsto\hat{U}\ket{\psi}$ and the corresponding nullifier transforms covariantly, $\hat{\delta}\mapsto\hat{U}\hat{\delta}\hat{U}^\dagger$. We will use this property to specify nullifiers of several CV states. Especially in our work, we consider CV states whose nullifiers are quadratic or lower in the creation and annihilation operators, and hence, can be stabilized with experimentally friendly conditional Gaussian operations.

\emph{Vacuum and squeezed vacuum state.} The vacuum state $\ket{0}$ is the ground state of the oscillator, whose nullifier is $\hat{a}$. The $\hat{p}$-squeezed vacuum state is obtained by the squeezing operator $\hat{S}(r)=\exp\big[r(\hat{a}^2-(\hat{a}^\dagger)^2)/2\big]$ with $r\in\mathbb{R}$, namely $\ket{p\simeq 0}=\hat{S}(-r)\ket{0}$. Its nullifier is
\begin{equation}
    \hat{S}(-r)\hat{a}\,\hat{S}^\dagger(-r)=\hat{a}\cosh r-\hat{a}^\dagger\sinh r.
    \label{eq:Nullifier_sq}
\end{equation}
In the limit $r\to\infty$, the nullifier in Eq.~\eqref{eq:Nullifier_sq} becomes proportional to the momentum operator $\hat{p}$. Correspondingly, $\ket{p\simeq0}$ approaches the momentum eigenstate $\ket{p=0}$, which satisfies $\hat{p}\ket{p=0}=0$.% Thus $\hat{p}$ is a nullifier of $\ket{p=0}$.

\emph{Cubic phase state.} The ideal cubic phase state is $\ket{\mathrm{CPS}}=\exp\big(i\eta\hat{x}^3/3\big)\ket{p=0}$. Experimentally, one uses the finite-squeezing version $\ket{\mathrm{CPS}^\ast}=\exp\big(i\eta\hat{x}^3/3\big)\ket{p\simeq0}$. Its nullifier is
\begin{align}
    \hat{\delta}_{\mathrm{CPS}}(r,\eta)&=\exp\!\left(i\tfrac{\eta}{3}\hat{x}^3\right)\big(\hat{a}\cosh r-\hat{a}^\dagger\sinh r\big)\exp\!\left(-i\tfrac{\eta}{3}\hat{x}^3\right)\nonumber\\
    &=\exp\!\left(i\tfrac{\eta}{3}\hat{x}^3\right)\left(i\tfrac{\mathrm{e}^{r}}{\sqrt{2}}\hat{p}+\tfrac{\mathrm{e}^{-r}}{\sqrt{2}}\hat{x}\right)\exp\!\left(-i\tfrac{\eta}{3}\hat{x}^3\right)\nonumber\\
    &= i\frac{\mathrm{e}^{r}}{\sqrt{2}}(\hat{p}-\eta\hat{x}^2)+\frac{\mathrm{e}^{-r}}{\sqrt{2}}\hat{x}.
    \label{eq:nullifier_CPS}
\end{align}
In the limit $r\to\infty$, the nullifier approaches $\hat{p}-\eta\hat{x}^2$, i.e., that of the ideal cubic phase state.

\emph{Trisqueezed state.} The trisqueezed state is $\ket{\mathrm{TSS}}=\hat{T}(\xi)\ket{0}$, where $\hat{T}(\xi)=\exp\big[\xi\{(\hat{a}^\dagger)^3-\hat{a}^3\}/3\big]$ denotes the trisqueezing operator and $\xi\in\mathbb{R}$ denotes the trisqueezing level. Its nullifier is
\begin{align}
    \hat{\delta}_{\mathrm{TSS}}(\xi)&=\hat{T}(\xi)\hat{a}\hat{T}^\dagger(\xi)\nonumber\\
    &=\hat{a}-\xi(\hat{a}^\dagger)^2+\mathcal{O}\left(\xi^2\right).
    \label{eq:stabilizer_TSS}
\end{align}
In this work, we consider the approximate trisqueezed state with sufficiently small trisqueezing level $\xi$ so that we can neglect $\mathcal{O}\left(\xi^2\right)$ terms.

\emph{Cat state and squeezed cat state.} The cat state is $\ket{\mathrm{CAT}_{\pm}}=\mathcal{N}_{\pm}(\alpha)(\ket{\alpha}\pm\ket{-\alpha})$, where $\ket{\alpha}=\hat{D}(\alpha)\ket{0}$ is a coherent state, $\hat{D}(\alpha)=\exp(\alpha\hat{a}^\dagger-\alpha^\ast\hat{a})$ is a displacement operator, $\alpha\in\mathbb{C}$ corresponds to the amplitude of the cat state, and $\mathcal{N}_{\pm}(\alpha)$ is a normalization factor. Its nullifier is
\begin{equation}
    \hat{\delta}_{\mathrm{CAT}}(\alpha)=\hat{a}^2-\alpha^2.
    \label{eq:stabilizer_CAT}
\end{equation}

In the same way, the squeezed cat state is $\ket{\mathrm{SqCAT}_{\pm}}=\mathcal{N}_{\pm}(\alpha, r)(\ket{\alpha, r}\pm\ket{-\alpha, r})$, where $\ket{\alpha, r}=\hat{D}(\alpha)\hat{S}(r)\ket{0}$ is a coherent squeezed state, and $\mathcal{N}_{\pm}(\alpha, r)$ is a normalization factor. The squeezed cat state can also be described as
\begin{align}
    \ket{\mathrm{SqCAT}_{\pm}}&=\mathcal{N}_{\pm}(\alpha, r)(\hat{D}(\alpha)\hat{S}(r)\ket{0}\pm\hat{D}(-\alpha)\hat{S}(r)\ket{0})\nonumber\\
    &=\mathcal{N}_{\pm}(\alpha, r)(\hat{S}(r)\hat{D}(\beta)\ket{0}\pm\hat{S}(r)\hat{D}(-\beta)\ket{0})\nonumber\\
    &=\hat{S}(r)\left[\mathcal{N}_{\pm}(\alpha, r)\left\{\hat{D}(\beta)\ket{0}\pm\hat{D}(-\beta)\ket{0}\right\}\right],
\end{align}
where $\beta=\alpha\cosh(r)+\alpha^\ast\sinh(r)$. Hence, the nullifier of the squeezed cat state is obtained from the nullifier of the cat state by conjugating the squeezing operator
\begin{align}
    \hat{\delta}_{\mathrm{SqCAT}}(\alpha, r)&=\hat{S}(r)\left(\hat{a}^2-\beta^2\right)\hat{S}^\dagger(r)\nonumber\\
    &=\left\{\hat{a}\cosh(r)+\hat{a}^\dagger\sinh(r)\right\}^2-\beta^2.
    \label{eq:stabilizer_SqCAT}
\end{align}

% -------------------------------------------------------
\section{Autonomous quantum error correction}
\label{sec:gate_level}

\begin{figure}[t]
    \centering
    \includegraphics[width=0.9\linewidth]{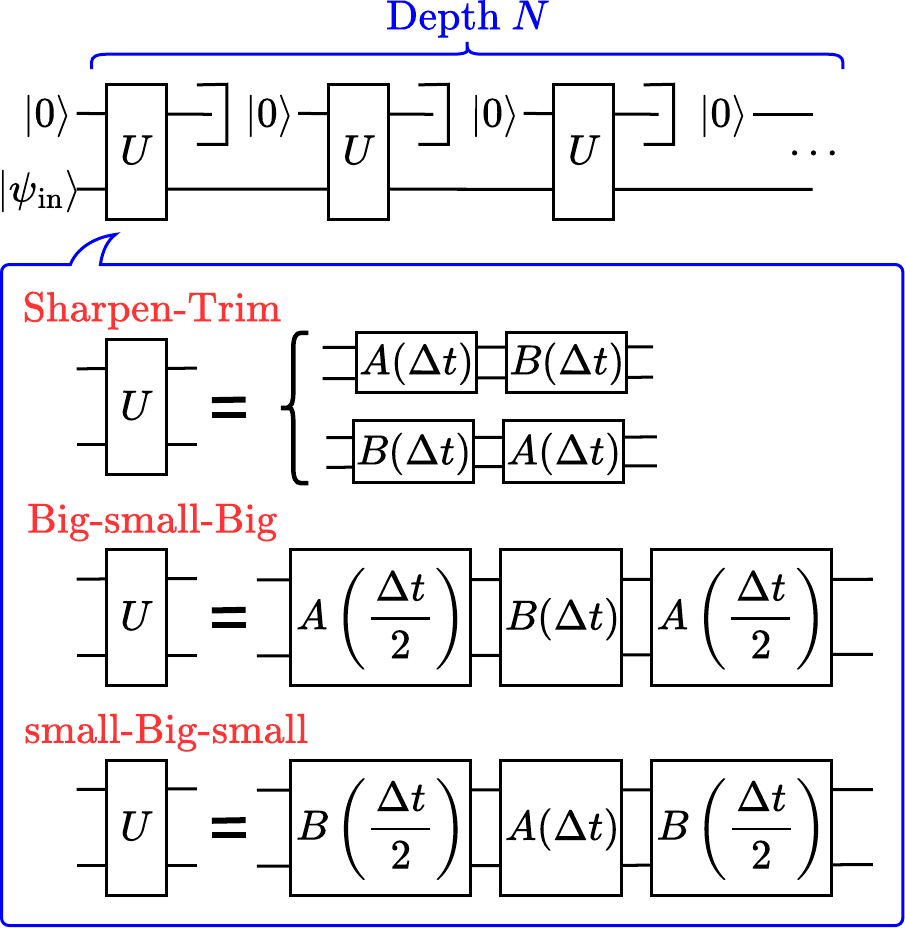}
    \caption{\textbf{Gate-level circuits for the autonomous QEC.} Define $\hat{A}(\Delta t)=\exp\!\left[-i\Gamma\Delta t\,\{\hat{X}\otimes(\hat{\delta}+\hat{\delta}^\dagger)/2\}\right]$ and $\hat{B}(\Delta t)=\exp\!\left[-i\Gamma\Delta t\,\{-\hat{Y}\otimes(\hat{\delta}-\hat{\delta}^\dagger)/(2i)\}\right]$. Repeated application of Trotterized products of $\hat{A}$ and $\hat{B}$ stabilizes the desired oscillator subspace.}
    \label{fig:Circuit_General}
\end{figure}

We review the autonomous QEC~\cite{PhysRevLett.125.260509} in Fig.~\ref{fig:Circuit_General}. Once a nullifier $\hat{\delta}$ of the target state is identified, the target can be prepared by coupling the oscillator to a bath such that the dynamics drives the oscillator toward a subspace spanned by states satisfying $\hat{\delta}\ket{\psi}=0$.

Consider the oscillator--bath interaction
\begin{equation}
    \hat{H}=\Gamma\big(\hat{a}^\dagger_{\mathrm{bath}}\otimes \hat{\delta}+\hat{a}_{\mathrm{bath}}\otimes \hat{\delta}^\dagger\big),
    \label{eq:HamilInit}
\end{equation}
where $\hat{a}_{\mathrm{bath}}$ is an annihilation operator acting on an ancilla bath, and $\Gamma$ is a coupling strength. We take the initial state to be $\ket{0}\!\bra{0}_{\mathrm{bath}}\otimes \hat{\rho}$. We then let the system evolve for a short time $\Delta t$ and trace out the bath. The reduced state is changed into
\begin{equation}
    \hat{\rho}+ (\Gamma\Delta t)^2\!\left(\hat{\delta}\hat{\rho}\hat{\delta}^\dagger-\tfrac{1}{2}\hat{\delta}^\dagger\hat{\delta}\hat{\rho}-\tfrac{1}{2}\hat{\rho}\hat{\delta}^\dagger\hat{\delta}\right)+\mathcal{O}(\Delta t^{3}),
    \label{eq:LindToDissipation}
\end{equation}
which approximates Gorini–Kossakowski–Sudarshan–Lindblad (GKSL) equation
\begin{equation}
    \frac{\mathrm{d}\hat{\rho}}{\mathrm{d}t}=\kappa \left(\hat{\delta}\hat{\rho}\hat{\delta}^\dagger-\frac{1}{2}\hat{\delta}^\dagger\hat{\delta}\hat{\rho}-\frac{1}{2}\hat{\rho}\hat{\delta}^\dagger\hat{\delta}\right),
    \label{eq:Lindblad}
\end{equation}
where $\kappa$ represents the dissipation rate. The steady states of Eq.~\eqref{eq:Lindblad} live in the subspace spanned by the states satisfying $\hat{\delta}\ket{\psi}=0$. By repeating the following procedure—preparing $\ket{0}_{\mathrm{bath}}$, evolving under $\hat{H}$, and resetting the bath—we relax the state $\hat{\rho}$ toward the target state.

To implement this at the gate level, we discretize the evolution and replace the bath by a qubit. Over short steps $\Delta t$ (such that the excitations of bath per step $\ll 1$), the bath can be replaced with an ancillary qubit $\hat{a}_{\mathrm{bath}}\rightarrow(\hat{X}- i\hat{Y})/2$, with Pauli operators $\hat{X},\hat{Y}$. Then, the unitary operation per step is
\begin{align}
    \hat{U}&=\exp\!\left[-i\Gamma\Delta t\big(\hat{a}^\dagger_{\mathrm{bath}}\otimes \hat{\delta}+\hat{a}_{\mathrm{bath}}\otimes \hat{\delta}^\dagger\big)\right]\nonumber\\
    &=\exp\!\left[-i\Gamma\Delta t\left(\hat{X}\otimes\frac{\hat{\delta}+\hat{\delta}^\dagger}{2}-\hat{Y}\otimes\frac{\hat{\delta}-\hat{\delta}^\dagger}{2i}\right)\right].
    \label{eq:UnitaryAll}
\end{align}
We approximate Eq.~\eqref{eq:UnitaryAll} using Trotter decompositions. Reference~\cite{PhysRevLett.125.260509} considers three types of decompositions:
\begin{align}
    \text{Sharpen--Trim:}\ &\hat{U}^{(\mathrm{ST})}=\begin{cases}
        \hat{A}(\Delta t)\hat{B}(\Delta t) \ \text{(Sharpen)}\\
        \hat{B}(\Delta t)\hat{A}(\Delta t) \ \text{(Trim)}
    \end{cases},\\
    \text{Big--small--Big:}\  &\hat{U}^{(\mathrm{BsB})}=\hat{A}\left(\frac{\Delta t}{2}\right)\hat{B}(\Delta t)\hat{A}\left(\frac{\Delta t}{2}\right),\\
    \text{small--Big--small:}\  &\hat{U}^{(\mathrm{sBs})}=\hat{B}\left(\frac{\Delta t}{2}\right)\hat{A}(\Delta t)\hat{B}\left(\frac{\Delta t}{2}\right),
    \label{eq:ST_BsB_sBs}
\end{align}
where
\begin{align}
    \hat{A}(\Delta t)&=\exp\!\left[-i\Gamma\Delta t\Big(\hat{X}\otimes\tfrac{\hat{\delta}+\hat{\delta}^\dagger}{2}\Big)\right],\\
    \hat{B}(\Delta t)&=\exp\!\left[-i\Gamma\Delta t\Big(-\hat{Y}\otimes\tfrac{\hat{\delta}-\hat{\delta}^\dagger}{2i}\Big)\right].
\end{align}
In this paper, we refer to $\hat{A}(\Delta t)$ as ``Big,'' and $\hat{B}(\Delta t)$ as ``small'' operator. The Sharpen--Trim unitary corresponds to the first-order Trotterization, while Big--small--Big and small--Big--small unitaries are of the second order. 
In the Sharpen--Trim protocol, we apply the Sharpen unitary operator and the Trim operator alternately. Experimentally, the Pauli operators acting on the ancillary qubit are realized by basis rotations followed by a $Z$-type conditional interaction. Specifically, the interaction generated by $\hat{X}\otimes \hat{O}$ is implemented by applying a Hadamard gate $\hat{H}$ to the ancillary qubit before and after the evolution, using $\hat{H}\hat{Z}\hat{H}=\hat{X}$. Similarly, the $\hat{Y}$ interaction is obtained via appropriate single-qubit rotations, e.g., $\exp (i\pi\hat{X}/4)\hat{Z}\exp (-i\pi\hat{X}/4)=\hat{Y}$. We can choose any one of the three decompositions to maximize the performance of the autonomous QEC, depending on the target state.

Hereafter, we detail stabilization of the cubic phase state, approximate trisqueezed state, cat state, and squeezed cat state using these constructions.

% -------------------------------------------------------
\section{Physical implementation}\label{sec:physicalimplementation}
In this section, we apply the autonomous QEC introduced in Sec.~\ref{sec:gate_level} to the CV quantum states introduced in Sec.~\ref{sec:cv_states}. To perform the autonomous QEC in a scalable and controllable manner, we decompose the protocol into a sequence of elementary gates, such as displacement, rotation, and squeezing, rather than directly constructing the corresponding Hamiltonian. The concrete decomposition is described below.

\subsection{Cubic phase state}
To stabilize a cubic phase state, we require the Hamiltonians
\begin{align}
    \hat{X}\otimes\big[\hat{\delta}_{\mathrm{CPS}}(r,\eta)+\hat{\delta}_{\mathrm{CPS}}^{\dagger}(r,\eta)\big] &= \hat{X}\otimes\big[\sqrt{2}\,\mathrm{e}^{-r}\hat{x}\big], \label{eq:Re_CPS}\\
    \hat{Y}\otimes\frac{\hat{\delta}_{\mathrm{CPS}}(r,\eta)-\hat{\delta}_{\mathrm{CPS}}^{\dagger}(r,\eta)}{i} &= \hat{Y}\otimes\big[\sqrt{2}\,\mathrm{e}^{r}(\hat{p}-\eta\hat{x}^2)\big], \label{eq:Im_CPS}
\end{align}
using Eq.~\eqref{eq:nullifier_CPS}. The term in Eq.~\eqref{eq:Re_CPS} corresponds to a conditional displacement operation. The term in Eq.~\eqref{eq:Im_CPS} can be decomposed into standard Gaussian operations as
\begin{align}
    &\exp\big[i(\hat{p}-\eta\hat{x}^2) t\big]=\nonumber\\
    &\quad \hat{D}\big(\alpha_{\mathrm{CPG}}(t)\big)\,\hat{R}\big(\phi_{\mathrm{CPG},2}(t)\big)\,\hat{S}\big(r_{\mathrm{CPG}}(t)\big)\,\hat{R}\big(\phi_{\mathrm{CPG},1}(t)\big),
    \label{eq:Gauss_decomposition}
\end{align}
where $\hat{R}(\phi)=\exp(-i\phi\hat{a}^\dagger\hat{a})$ is the rotation, with $t,\phi\in\mathbb{R}$ and $\alpha\in\mathbb{C}$, and
\begin{equation}
\begin{aligned}
    \alpha_{\mathrm{CPG}}(t)&=\frac{-t+i\eta t^2}{\sqrt{2}},\\
    \phi_{\mathrm{CPG},1}(t)&=\arctan\big(\sqrt{1+\eta^2 t^2}-\eta t\big),\\
    r_{\mathrm{CPG}}(t)&=\ln\big(\sqrt{1+\eta^2 t^2}-\eta t\big),\\
    \phi_{\mathrm{CPG},2}(t)&=-\arctan\big(\sqrt{1+\eta^2 t^2}+\eta t\big).
    \label{eq:EvolutionTime}
\end{aligned}
\end{equation}
(see Appendix~\ref{sppl:Bloch-messiah} for its proof). Thus, realizing the cubic phase dissipator requires conditional rotation, squeezing, and displacement operations. Conditional rotation and displacement have been demonstrated on multiple platforms~\cite{sivak2023real, eickbusch2022fast}. Conditional squeezing has been proposed in Refs.~\cite{ayyash2024driven, del2024controlled}; although its direct implementation is not experimentally demonstrated, it can be further decomposed into conditional rotations and single-mode squeezing~\cite{q15h-nmq9}, supporting the feasibility of our scheme.

\subsection{Trisqueezed state}

To stabilize a trisqueezed state, we require the Hamiltonians
\begin{align}
    &\hat{X}\otimes\big[\hat{\delta}_{\mathrm{TSS}}(\xi)+\hat{\delta}_{\mathrm{TSS}}^{\dagger}(\xi)\big] \nonumber\\
    &= \hat{X}\otimes\big[\sqrt{2}\hat{x}-\xi(\hat{x}^2-\hat{p}^2)+\mathcal{O}(\xi^2)\big], \label{eq:Re_TSS}\\
    &\hat{Y}\otimes\frac{\hat{\delta}_{\mathrm{TSS}}(\xi)-\hat{\delta}_{\mathrm{TSS}}^{\dagger}(\xi)}{i} \nonumber\\
    &= \hat{Y}\otimes\big[\sqrt{2}\hat{p}+\xi(\hat{x}\hat{p}+\hat{p}\hat{x})+\mathcal{O}(\xi^2)\big]. \label{eq:Im_TSS}
\end{align}
Since Hamiltonians containing third or higher-order terms in the quadrature operators $\hat{x}$ and $\hat{p}$ are experimentally challenging to implement, we restrict our consideration to the regime that can be realized with second-order Hamiltonians: Gaussian operations. Accordingly, $\mathcal{O}(\xi^2)$ terms are neglected in the following analysis.% This restriction induces the error when $\xi$ is large, and such an error is numerically analyzed in Sec.~\ref{sec:numerics}.

The term in Eq.~\eqref{eq:Re_TSS} can be decomposed into standard Gaussian operations as
\begin{align}
    &\exp\left[-i\left\{\sqrt{2}\hat{x}-\xi(\hat{x}^2-\hat{p}^2)\right\}t\right]\nonumber\\
    &\quad =\hat{D}\left(\frac{1-\cosh(2\xi t)-i\sinh(2\xi t)}{2\xi}\right)\nonumber\\
    &\quad\times\hat{R}\left(\frac{\pi}{4}\right)\hat{S}\big(-2\xi t\big)\hat{R}\left(-\frac{\pi}{4}\right).
    \label{eq:Gauss_decomposition_TSS}
\end{align}
In the same way, the term in Eq.~\eqref{eq:Im_TSS} can be decomposed into other standard Gaussian operations as
\begin{align}
    &\exp\left[-i\left\{\sqrt{2}\hat{p}+\xi(\hat{x}\hat{p}+\hat{p}\hat{x})\right\}t\right]\nonumber\\
    &\quad =\hat{D}\left(\frac{\exp(2\xi t)-1}{2\xi}\right)\hat{S}(-2\xi t).
    \label{eq:Gauss_decomposition_TSS_2}
\end{align}
(see Appendix~\ref{sppl:Bloch-messiah} for their proofs). Hence, we can realize the trisqueezing dissipator with conditional Gaussian operations, which support the feasibility of our scheme.

\subsection{Cat state and squeezed cat state}

To stabilize a cat state, we require the Hamiltonians
\begin{align}
    &\hat{X}\otimes\big[\hat{\delta}_{\mathrm{CAT}}(\alpha)+\hat{\delta}_{\mathrm{CAT}}^{\dagger}(\alpha)\big] \nonumber\\
    &= \hat{X}\otimes\big[(\hat{x}^2-x_0^2) - (\hat{p}^2-p_0^2)\big], \label{eq:Re_CAT}\\
    &\hat{Y}\otimes\frac{\hat{\delta}_{\mathrm{CAT}}(\alpha)-\hat{\delta}_{\mathrm{CAT}}^{\dagger}(\alpha)}{i} \nonumber\\
    &= \hat{Y}\otimes\big[\hat{x}\hat{p}+\hat{p}\hat{x}-2x_0p_0\big], \label{eq:Im_CAT}
\end{align}
with $\alpha=(x_0+ip_0)/\sqrt{2},\ x_0, p_0\in\mathbb{R}$. The Hamiltonians $\hat{x}^2-\hat{p}^2$ and $\hat{x}\hat{p}+\hat{p}\hat{x}$ in Eqs.~\eqref{eq:Re_CAT} and~\eqref{eq:Im_CAT} correspond to the $p$-squeezing operation and $x$-squeezing operation, respectively. Hence, we can realize the cat-state dissipator with conditional squeezing operations.

In the same way, to stabilize a squeezed cat state, we require the Hamiltonians
\begin{align}
    &\hat{X}\otimes\big[\hat{\delta}_{\mathrm{SqCAT}}(\alpha,r)+\hat{\delta}_{\mathrm{SqCAT}}^{\dagger}(\alpha,r)\big] \nonumber\\
    &= \hat{X}\otimes\big[(\hat{x}^2-x_0^2)\mathrm{e}^{2r} - (\hat{p}^2-p_0^2)\mathrm{e}^{-2r}\big], \label{eq:Re_SqCAT}\\
    &\hat{Y}\otimes\frac{\hat{\delta}_{\mathrm{SqCAT}}(\alpha,r)-\hat{\delta}_{\mathrm{SqCAT}}^{\dagger}(\alpha,r)}{i} \nonumber\\
    &= \hat{Y}\otimes\big[\hat{x}\hat{p}+\hat{p}\hat{x}-2x_0p_0\big].
    \label{eq:Im_SqCAT}
\end{align}
The CV-system term in Eq.~\eqref{eq:Re_SqCAT} can be decomposed into the rotation and squeezing operators as
\begin{align}
    \exp&\left[-i(\hat{x}^2\mathrm{e}^{2r}-\hat{p}^2\mathrm{e}^{-2r})t\right]\nonumber\\
    &=\hat{R}\left(\phi_{\mathrm{SqCAT}}(t)+\frac{\pi}{2}\right)\,\hat{S}\big(r_{\mathrm{SqCAT}}(t)\big)\,\hat{R}\left(\phi_{\mathrm{SqCAT}}(t)\right),
    \label{eq:Gauss_decomposition_SqCAT}
\end{align}
with
\begin{align}
    \tanh(r_{\mathrm{SqCAT}}(t))&=-\frac{\cosh(2r)\sinh(2t)}{\sqrt{1+\cosh^2(2r)\sinh^2(2t)}}\\
    \tan\left(2\phi_{\mathrm{SqCAT}}(t)\right)&=\frac{\cosh(2t)}{\sinh(2r)\sinh(2t)}.
\end{align}
The term in Eq.~\eqref{eq:Im_SqCAT} can be implemented with the conditional squeezing operation, as is the case with Eq.~\eqref{eq:Im_CAT} (see Appendix~\ref{sppl:Bloch-messiah} for their proofs).

\section{Use cases and their efficiency}\label{Sec:UseCasesAnd}
We analyze the efficiency of our algorithm in two use cases: (1) preparing a target quantum state and (2) suppressing an error of a state in a code space. For state preparation, we evaluate the depth of the autonomous QEC circuit required to reach high fidelity. For error suppression, we evaluate the leakage from the code space in the steady state under a given strength of noise and a rate of QEC cycles.
%For error suppression, we evaluate the coupling strength $\Gamma$ needed to suppress error.

For the state-preparation case, we estimate the relaxation time to the ground state of the GKSL dynamics in Eq.~\eqref{eq:Lindblad} by working in an appropriate basis. Suppose we aim to prepare a state $\ket{\psi}$ with nullifier $\hat{\delta}$, and we have a basis $\{\ket{\tilde{n}} \mid \tilde{n}\in\mathbb{Z}_{\ge 0}\}$ whose ground state coincides with the target, $\ket{\tilde{0}}=\ket{\psi}$, and whose annihilation operator is the nullifier, $\tilde{a}=\hat{\delta}$. For example, if $\ket{\psi}$ is produced from the vacuum state by a unitary operation $\hat{U}$ as $\ket{\psi}=\hat{U}\ket{0}$, we take $\ket{\tilde{n}}=\hat{U}\ket{n}$ and $\tilde{a}=\hat{U}\hat{a}\hat{U}^\dagger$, where $\{\ket{n}\}$ is the Fock basis and $\hat{a}$ its annihilation operator. We then have:

\textbf{Theorem 1 (Preparing a quantum state).}
Let $\ket{\phi}$ be the initial state and $\ket{\psi}$ the target state. Let $\{\ket{\tilde{n}}\}$ be a basis with $\ket{\tilde{0}}=\ket{\psi}$ and annihilation operator $\tilde{a}=\hat{\delta}$. Evolve $\ket{\phi}$ under the GKSL dynamics in Eq.~\eqref{eq:Lindblad} for time $\tau$ to obtain $\hat{\rho}(\tau)$. We define a fidelity between quantum states $F(\hat{\rho},\hat{\sigma}):=\mathrm{Tr}\left(\sqrt{\sqrt{\hat{\rho}}\hat{\sigma}\sqrt{\hat{\rho}}}\right)$. To achieve the fidelity $F(\hat{\rho}(\tau),\ket{\psi}\bra{\psi})=1-\epsilon/2$, the required duration is estimated as
\[
\kappa\tau \simeq \log\!\left(\frac{\langle \tilde{n} \rangle}{\epsilon}\right),
\quad
\text{where }
\langle \tilde{n} \rangle=\bra{\phi}\tilde{a}^\dagger \tilde{a}\ket{\phi}.
\]
If we approximate the GKSL dynamics via the oscillator--bath map in Eq.~\eqref{eq:LindToDissipation}, the duration satisfies $\kappa\tau \approx N(\Gamma\Delta t)^2$, where $N$ is the circuit depth, $\Gamma$ the coupling strength, and $\Delta t$ the short evolution time. Hence the depth required to reach fidelity $1-\epsilon/2$ is approximated as
\begin{equation}
    N \simeq \frac{1}{(\Gamma\Delta t)^2}\,
    \log\!\left(\frac{\langle \tilde{n} \rangle}{\epsilon}\right).
    \label{eq:DepthTheory}
\end{equation}
See Appendix~\ref{app:ProofsOfEfficiency} for its proof. We note that the Trotterization error is not accounted for in Theorem~1. Instead, it is numerically evaluated in Sec.~\ref{sec:numerics}.

Hereafter, we apply Theorem~1 to specific target states, namely the cubic phase state and the trisqueezed state prepared from the vacuum state. For the cubic phase state, the excitation number of the initial vacuum state, $\langle \tilde{n}\rangle_{\mathrm{CPS}}
=\bra{0}\hat{\delta}^\dagger_{\mathrm{CPS}}(r,\eta)\,\hat{\delta}_{\mathrm{CPS}}(r,\eta)\ket{0}$,
can be evaluated as
\begin{equation}
    \langle \tilde{n}\rangle_{\mathrm{CPS}}
    =\frac{1}{2}\cosh(2r)+\frac{3}{8}\eta^{2}\mathrm{e}^{2r}-\frac{1}{2},
    \label{eq:ExcitationCPS}
\end{equation}
see Appendix~\ref{app:ProofsOfEfficiency} for the derivation. Substituting Eq.~\eqref{eq:ExcitationCPS} into Eq.~\eqref{eq:DepthTheory}, we obtain an estimate of the circuit depth $N$ required to prepare the cubic phase state from the vacuum state. In particular, in the large-squeezing regime $r\gg 1$, the required depth is approximated as
\begin{equation}
    N=\frac{1}{(\Gamma\Delta t)^2}
    \left[
        2r+\log\!\left(\frac{3\eta^{2}+2}{8\epsilon}\right)
    \right],
\end{equation}
which exhibits linear scaling in the squeezing parameter $r$ and logarithmic scaling in the cubicity parameter $\eta$, leading to an efficient preparation of the cubic phase state with high squeezing and high cubicity. 

In the same manner, we also estimate the required depth for the trisqueezed-state preparation. The excitation number $\langle \tilde{n}\rangle_{\mathrm{TSS}}=\bra{0}\hat{\delta}^{\dagger}_{\mathrm{TSS}}(\xi)\,\hat{\delta}_{\mathrm{TSS}}(\xi)\ket{0}$ can be evaluated as
\begin{equation}
    \langle \tilde{n}\rangle_{\mathrm{TSS}}=2\xi^2+\mathcal{O}(\xi^3),
    \label{eq:ExcitationTSS}
\end{equation}
see Appendix~\ref{app:ProofsOfEfficiency} for the derivation.  Substituting Eq.~\eqref{eq:ExcitationTSS} into Eq.~\eqref{eq:DepthTheory}, we obtain an estimate of the circuit depth $N$ required to prepare the trisqueezed state from the vacuum state.

For the error-suppression case, we estimate the performance of the proposed QEC protocol under a given strength of noise. As the figure of merit for the QEC protocol, we adopt leakage, or the population outside the target subspace.  It is defined as
\begin{align}
    w(\hat{\rho}_{\rm ss})=\Tr[(\hat{I}-\hat{P}_{\rm code})\hat{\rho}_{\rm ss}],
    \label{eq:leakage_def}
\end{align}
where $\hat{\rho}_{\rm ss}$ is the steady state under error and continual applications of the QEC circuit, and $\hat{P}_{\rm code}$ is the projector onto the target subspace.
To obtain a rough estimate of $w(\hat{\rho}_{\rm ss})$, we recall that the QEC protocol is designed to mimic the time evolution described by the GKSL equation in Eq.~\eqref{eq:Lindblad}.
When the QEC circuit is applied at rate $R$ in the presence of an error $\mathcal D[\hat{b}_{\rm n}]$ with rate $\kappa_{\rm n}$, the overall dynamics can be approximated by the continuous-time master equation
\begin{align}
    \od{1}{\hat{\rho}}{t} \simeq \kappa_{\rm s}{\mathcal D}[\hat{\delta}]\hat{\rho}+\kappa_{\rm n}{\mathcal D}[\hat{b}_{\rm n}]\hat{\rho}
    \label{eq:approxLindblad}
\end{align}
with $\kappa_{\rm s}=R(\Gamma\Delta t)^2$ being the effective stabilization rate.
To sufficiently suppress the effect of the noise in the steady state, the stabilization rate should be much larger than the noise ratio:
\begin{align}
    \epsilon=\frac{\kappa_{\rm n}}{\kappa_{\rm s}}=\frac{\kappa_{\rm n}}{R(\Gamma\Delta t)^2}\ll 1.
    \label{eq:epsilon}
\end{align}
In this regime, the leakage admits the expansion
\begin{align}
    w(\hat{\rho}_{\rm ss})=C+\epsilon A + O(\epsilon^2).
    \label{eq:leakageExpression}
\end{align}
Here, the first term $C(>0)$ represents the leakage present even in the absence of noise, originating from the Trotterization error in the QEC circuit, while the second term represents the leading-order contribution from the noise term.
The coefficient $A$ can be calculated using the perturbation theory for the master equation. We expand the steady state as $\hat{\rho}_{\rm ss}=\hat{\rho}_0+\epsilon \hat{\rho}_1+O(\epsilon^2)$, and substitute it to the steady state condition
\begin{align}
     {\mathcal D}[\hat{\delta}]\hat{\rho}_{\rm ss}+\epsilon{\mathcal D}[\hat{b}_{\rm n}]\hat{\rho}_{\rm ss}=0.
     \label{eq:steady state}
\end{align}
By solving Eq.~\eqref{eq:steady state} order by order, we obtain the first order correction to the leakage as $A=-\Tr[\hat{\rho}_1 \hat{P}_{\rm code}]$.
See Appendix~\ref{app:perturb-sqcat} for the details of the perturbative expansion of the steady state.
In Sec.~\ref{sec:numerics}, we numerically calculate the leakage for the squeezed cat code and confirm that the behavior of leakage is consistent with the expansion in~\eqref{eq:leakageExpression}, and the coefficient $A$ agrees well with the result from the perturbation theory.

\section{Numerical simulations}
\label{sec:numerics}

We perform numerical simulations for two use cases: (i) preparing a target quantum state, shown in Figs.~\ref{fig:QEC_generating} and~\ref{fig:QEC_generating_SqLevel}, and (ii) suppressing an error of a target quantum state, shown in Figs.~\ref{fig:QEC_protecting} and~\ref{fig:QEC_Leakage}. In the following, we describe the simulation setup and results for each case.

\subsection{Numerical simulations for preparing target quantum states}

\begin{figure}[t]
    \centering
    \includegraphics[width=1.0\linewidth]{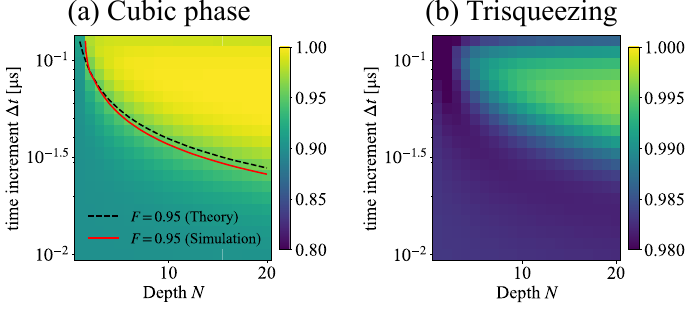}
    \caption{\textbf{Numerical simulations of preparing quantum states.} We prepare various states from the vacuum state with our autonomous QEC. The horizontal axis represents the circuit depth $N$, the vertical axis represents the time increment $\Delta t$, and the color scale indicates the fidelity with respect to the corresponding target state. (a) Cubic phase state with squeezing level $5\ \mathrm{dB}$ and $\eta=0.3$. (b) Trisqueezed state with trisqueezing level $2\ \mathrm{dB}$.
    }
    \label{fig:QEC_generating}
\end{figure}

\begin{figure}[t]
    \centering
    \includegraphics[width=1.0\linewidth]{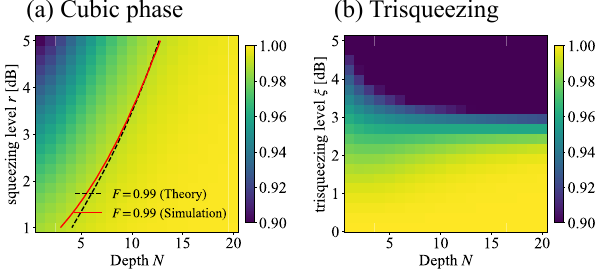}
    \caption{\textbf{Numerical simulations of preparing quantum states with different squeezing or trisqueezing levels.} The vertical axis is the squeezing or trisqueezing level instead of the time increment $\Delta t$. Here, $\Delta t$ is fixed at 50 ns; all other conditions are the same as those in Fig.~\ref{fig:QEC_generating}.
    }
    \label{fig:QEC_generating_SqLevel}
\end{figure}

We first consider the preparation of target quantum states in Fig.~\ref{fig:QEC_generating} to evaluate how the complexity of a preparable state depends on the number of preparation steps. Specifically, we study the cubic phase state and the trisqueezed state introduced in Sec.~\ref{sec:cv_states}. Figures~\ref{fig:QEC_generating}(a) and \ref{fig:QEC_generating}(b) show the results for the cubic phase and trisqueezed states, respectively. In these figures, the horizontal axis represents the circuit depth $N$, the vertical axis represents the time increment $\Delta t$, and the color scale indicates the fidelity with respect to the corresponding target state. Unless otherwise stated, other parameters such as the squeezing level $r$ are fixed. In each panel, the autonomous QEC is applied to the vacuum state as an initial state to prepare the target state. The solid and dashed curves in Fig.~\ref{fig:QEC_generating}(a) indicate the contour at fidelity $0.95$ obtained from the numerical simulation and the theoretical estimate, respectively. {We note that we do not plot such curves in Fig.~\ref{fig:QEC_generating}(b) because the fidelity contours of the trisqueezed state do not show good agreement with the theoretical estimates, which is attributed to the truncation of the $\mathcal{O}(\xi^2)$ terms in Eq.~\eqref{eq:stabilizer_TSS}.} The theoretical contour of the fidelity is estimated from Eq.~\eqref{eq:DepthTheory}. Figure~\ref{fig:QEC_generating_SqLevel} shows an alternative representation in which the vertical axis is the squeezing or trisqueezing level instead of the time increment $\Delta t$; all other conditions are the same as those in Fig.~\ref{fig:QEC_generating}. We plot Fig.~\ref{fig:QEC_generating_SqLevel} to explicitly show that, for the cubic phase state, the required depth $N$ grows only linearly with the squeezing level $r$, as predicted by Eq.~\eqref{eq:DepthTheory}. For the trisqueezed state, we plot it to show that the fidelity remains high only in the regime of small squeezing $\xi$, reflecting the approximation in Eq.~\eqref{eq:stabilizer_TSS}.

We now describe the detailed simulation conditions used in Figs.~\ref{fig:QEC_generating} and \ref{fig:QEC_generating_SqLevel}. For Fig.~\ref{fig:QEC_generating}, we prepare a cubic phase state with squeezing level $5\ \mathrm{dB}$ and $\eta=0.3$, and a trisqueezed state with trisqueezing level $2\ \mathrm{dB}$. We employ the small--Big--small protocol, which is numerically verified to be the most effective among the three protocols for the cubic phase state and the trisqueezed state (see Appendix~\ref{app:NumericalVerificationOf} for details). The autonomous QEC is implemented with a coupling strength $\Gamma=10\ \mathrm{MHz}$, comparable to that used in previous experiments~\cite{eriksson2024universal}. The circuit depth is sampled as $N=1,2,\ldots,20$, while the time increment $\Delta t$ is sampled from $10\ \mathrm{ns}$ to $130\ \mathrm{ns}$. To incorporate experimental imperfections, we include ancillary-qubit decoherence with $T_1=T_2=100\ \mathrm{\mu s}$ and photon-loss errors with a damping rate of $5\ \mathrm{kHz}$, which is the same order as the previous experiments~\cite{Ni2023BeatingBreakEven, CampagneIbarcq2020, sivak2023real}. In Fig.~\ref{fig:QEC_generating_SqLevel}, the vertical axis corresponds to the squeezing or trisqueezing level, sampled from $1\ \mathrm{dB}$ to $5\ \mathrm{dB}$, while the time increment $\Delta t$ is fixed at $50\ \mathrm{ns}$.

From Figs.~\ref{fig:QEC_generating} and \ref{fig:QEC_generating_SqLevel}, we find that for both the cubic phase and trisqueezed states, the achievable fidelity remains close to unity even in the presence of the modeled imperfections if we choose parameters correctly, indicating that the proposed protocol can reliably prepare these non-Gaussian states. For the cubic phase state, the fidelity contours obtained from numerical simulations show good agreement with the theoretical estimates. We further observe that the required depth $N$ for the cubic phase state increases approximately linearly with the squeezing level $r$ in Fig.~\ref{fig:QEC_generating_SqLevel}(a), consistent with the scaling predicted by Eq.~\eqref{eq:DepthTheory}. This behavior demonstrates the effectiveness and scalability of our scheme. We can also create the trisqueezed state with trisqueezing level roughly up to $2\ \mathrm{dB}$ in Fig.~\ref{fig:QEC_generating_SqLevel}(b), indicating the limitation due to the approximation of $\mathcal{O}(\xi^2)$ term in Eq.~\eqref{eq:stabilizer_TSS}.

\subsection{Numerical simulations for suppressing an error}

\begin{figure}[t]
    \centering
    \includegraphics[width=1.0\linewidth]{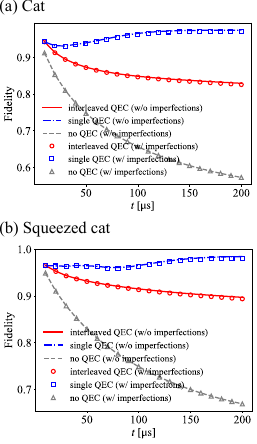}
    \caption{\textbf{Numerical simulations of suppressing the error of quantum states.}
    Time evolution of the fidelity between the ideal target state and the noisy state under three strategies:
    no error correction (``no QEC''), a single application of the autonomous QEC immediately before readout (``single QEC''), and repeated applications during storage (``interleaved QEC''). Here, we consider the dephasing error as the target error during the storage. Lines show the idealized numerical results without imperfections during the QEC operations, whereas points show the results including such imperfections. (a) Cat state with the coherent-state amplitude $\alpha = 3$. (b) Squeezed cat state with the amplitude $\alpha = 3$ and squeezing level $5~\mathrm{dB}$.
    }
    \label{fig:QEC_protecting}
\end{figure}

For the second use case of suppressing an error of a target quantum state, we numerically simulate the error suppression of the minus cat state $\ket{\mathrm{CAT}_-}$, and the minus squeezed cat state $\ket{\mathrm{SqCAT}_-}$ in Fig.~\ref{fig:QEC_protecting}. In each panel, the horizontal axis represents the storage time under a fixed error process, and the vertical axis represents the fidelity between the ideal target state and the corresponding simulated state. The curve labeled ``no QEC'' corresponds to the result with the evolution under the error channel only. The curve labeled ``single QEC'' corresponds to the result with the evolution under the same error channel up to time $t$, followed by a single execution of our autonomous QEC just before evaluating the fidelity. The curve labeled ``interleaved QEC'' corresponds to the result with the repeated executions of the autonomous QEC at regular intervals during the storage, interleaved with the noisy evolution. Panels (a) and (b) show the behavior for the cat state and the squeezed cat state, respectively.

We now describe the simulation model and parameters used in Fig.~\ref{fig:QEC_protecting}. As target states, we consider (a) a minus cat state with the coherent-state amplitude $\alpha = 3$, and (b) a squeezed minus cat state with the amplitude $\alpha = 3$ and squeezing level $5~\mathrm{dB}$. The noisy dynamics are modeled by a GKSL equation,
\begin{equation}
    \frac{d\hat{\rho}}{dt}
    = \kappa \,\mathcal{D}[\hat{L}]\hat{\rho},
    \quad
    \mathcal{D}[\hat{L}]\hat{\rho}
    = \hat{L}\hat{\rho}\hat{L}^\dagger
      - \tfrac{1}{2}\{\hat{L}^\dagger \hat{L}, \hat{\rho}\},
      \label{eq:GKSL}
\end{equation}
where $\kappa$ is the error rate. Especially in our numerical simulation, we take the dephasing error with $\hat{L}=\hat{a}^\dagger\hat{a}$, which is known to be corrected with the cat and the squeezed cat code~\cite{Mirrahimi_2014, PhysRevA.106.022431}, as the target error to be corrected. In both cases, we set $\kappa = 5~\mathrm{kHz}$, which is comparable to experimentally relevant decoherence rates~\cite{Hu2019, sivak2023real}. For comparison, we perform simulations both with and without imperfections during the autonomous QEC. In the simulations including such imperfections, we take into account $T_1$ and $T_2$ errors for the ancillary qubit system, in addition to the CV mode error described above. $T_1$ and $T_2$ time is set to $100~\mathrm{\mu s}$, which is the same order as Refs.~\cite{Ni2023BeatingBreakEven, CampagneIbarcq2020, sivak2023real}. We choose the Big--small--Big protocol, which is numerically validated to be the most effective of the three methods for the cat and the squeezed cat states (see Appendix~\ref{app:NumericalVerificationOf} for the detailed comparison). We assume its gate speed $\Gamma$ of $10~\mathrm{MHz}$, consistent with recent experimental demonstrations of conditional Gaussian operations~\cite{sivak2023real, eickbusch2022fast}. For the ``interleaved QEC'' protocol, the QEC protocol is applied every 10 $\mathrm{\mu s}$ over a total duration of 200 $\mathrm{\mu s}$, resulting in 20 rounds. Each QEC execution consists of depth $N=1$ with a time increment $\Delta t = 0.01~\mathrm{\mu s}$. In contrast, for the ``single QEC'' protocol, we apply the QEC protocol only once, immediately before the readout, with the same time increment $\Delta t = 0.01~\mathrm{\mu s}$. At each readout time, the depth $N$ is set equal to the number of QEC rounds that would have been applied in the ``interleaved QEC'' protocol up to that time, so that the total QEC depth is the same in the two strategies. The parameters $N$ and $\Delta t$ are selected by scanning them and adopting the combination that yields high fidelity with low depth (see Appendix~\ref{app:NumericalVerificationOf}).

Figure~\ref{fig:QEC_protecting} shows that for both the cat and squeezed cat states, the proposed autonomous QEC suppresses the dephasing error of the non-Gaussian states. For each state, the ``single QEC'' curve lies above the ``no QEC'' curve, confirming that even a single application of our protocol prior to readout reduces the accumulated error. Interestingly, this ``single QEC'' strategy further enhances the performance compared to the ``interleaved QEC.'' This behavior can be understood by using the idea of the subsystem decomposition~\cite{Xu2023, PhysRevLett.128.110502} as follows. The dephasing errors acting on the cat code do not induce logical errors; instead, the dephasing error causes only excitations in the gauge subsystem. Therefore, the role of the QEC protocol is simply to return the gauge excitations to the code space. When QEC is applied frequently, the protocol acts on a weakly excited gauge state. In contrast, performing QEC only once at the end allows gauge excitations to accumulate during the error process, which are then removed collectively. Since the error-correction protocol includes a two-photon loss process acting on the gauge mode, its efficiency increases with the excitation number. As a result, the single QEC strategy more effectively dissipates accumulated gauge excitations, leading to improved overall performance. It is also noteworthy that the performance of the autonomous QEC is only weakly affected by experimental imperfections, because the protocol is executed on a timescale much shorter than that of the relevant error processes.

\begin{figure}[t]
    \centering
    \includegraphics[width=0.9\linewidth]{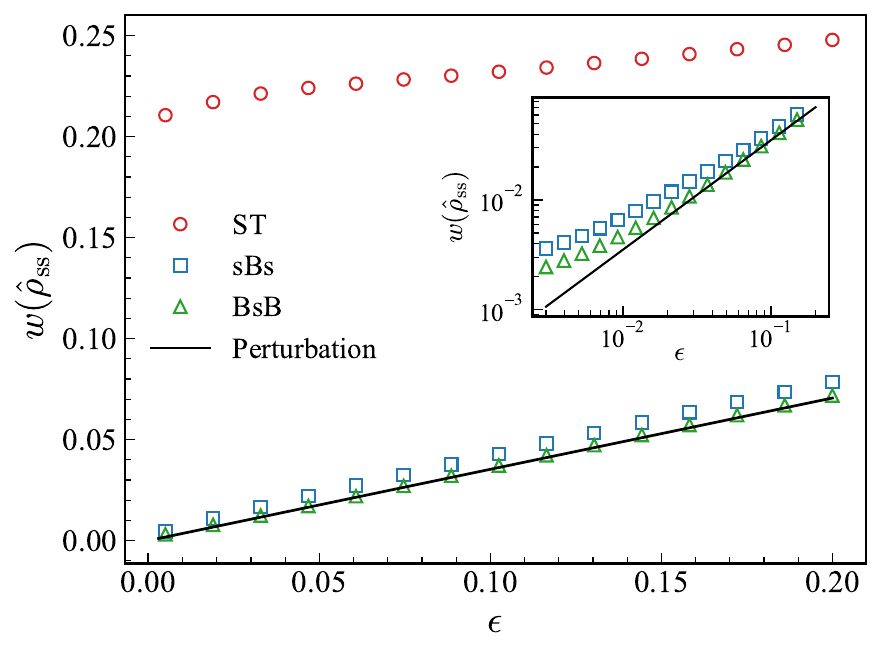}
    \caption{\textbf{Suppression of dephasing errors in the squeezed cat state.} Leakage is shown as a function of $\epsilon$ (Eq.~\eqref{eq:epsilon}) for different Trotterization schemes. The circles (``ST''), squares (``sBs''), and triangles (``BsB'') represent the numerical data obtained from the master equation, while the solid lines represent the contribution from the noise calculated using the perturbation theory. The inset shows a magnified view of the small-$\epsilon$ region. The parameters are $\alpha=1$, $r=0.5$, and $\Gamma \Delta t=0.13$. Here, ST, sBs, and BsB denote the Sharpen--Trim, small--Big--small, and Big--small--Big schemes, respectively.}
    \label{fig:QEC_Leakage}
\end{figure}

We also numerically verify the phenomenological expansion of the leakage in Eq.~\eqref{eq:leakageExpression} for the squeezed cat code.
%We also consider the suppression of the dephasing error with $\hat{b}_{\rm n}=\hat{a}^\dagger\hat{a}$ of the even squeezed cat state $\ket{\mathrm{SqCAT}_{+}}$.
In Fig.~\ref{fig:QEC_Leakage}, we numerically evaluate leakage in the steady state for different types of Trotterization, i.e., Sharpen--Trim, small--Big--small, and Big--small--Big. The horizontal axis denotes the noise ratio $\epsilon$ defined in Eq.~\eqref{eq:epsilon}, and the vertical axis denotes the leakage defined in Eq.~\eqref{eq:leakage_def}. The points correspond to the steady state obtained by numerically simulating the long-time evolution governed by Eq.~\eqref{eq:approxLindblad}. The lines show the contribution from the noise obtained from the perturbative analysis, which is detailed in Appendix~\ref{app:perturb-sqcat}.
To reduce the computational cost, we do not include imperfections during the QEC operations. However, as suggested by the results in Fig.~\ref{fig:QEC_protecting}, their effect is expected to be minor, so we expect that this simplification does not change the qualitative conclusion. From the inset in Fig.~\ref{fig:QEC_Leakage}, we observe a residual leakage ($C$ in Eq.~\eqref{eq:leakageExpression}) already at $\epsilon=0$, whose magnitude depends on the type of Trotterization.
Since Sharpen--Trim corresponds to the first-order Trotter decomposition, while small--Big--small and Big--small--Big correspond to the second-order, small--Big--small and Big--small--Big exhibit smaller residual leakage.
As the error strength is increased, the leakage increases according to the first-order perturbation in $\epsilon$, and this leading-order dependence is essentially independent of the Trotterization type and coincides with the result from the perturbation theory.

In summary, these results indicate that the autonomous QEC can suppress the error of the cat and squeezed cat states, supporting the practicality of our scheme.

% -------------------------------------------------------
\section{Discussion and conclusion}\label{sec:Discussion}
In this paper, we have provided a nullifier-based systematic construction of digital autonomous QEC. We have focused on two use cases: (i) state preparation and (ii) error suppression. For the state preparation, we have studied the preparation of two representative non-Gaussian resource states, the cubic phase state and the trisqueezed state, starting from the vacuum state. These states are key ingredients for universal CV quantum computation beyond what is efficiently simulatable with a classical computer. For the error suppression, we have considered the dephasing-error suppression of the cat states and squeezed cat states, which are promising bosonic QEC code states. All protocols proposed here are implementable using experimentally accessible conditional Gaussian operations.

We have also provided numerical evidence for the performance of our protocol. We have simulated both the state-preparation and error-suppression scenarios while incorporating realistic experimental imperfections, including photon loss and dephasing error in the bosonic mode, and qubit decoherence characterized by finite $T_1$ and $T_2$ times. For state preparation, we have analyzed the circuit depth $N$ and the time step $\Delta t$ required to prepare the cubic phase state and the trisqueezed state from the vacuum state. We have further examined the achievable squeezing or trisqueezing level: for the cubic phase state, the squeezing level increases approximately linearly with $N$, while for the trisqueezed state, we can prepare it with a moderate trisqueezing level of about $2\ \mathrm{dB}$. For error suppression, we have applied our autonomous QEC to the cat and the squeezed cat states subjected to dephasing error. We have found that implementing our protocol improves the fidelity compared with the passive evolution without QEC. Notably, applying the protocol once immediately before readout yields a larger fidelity recovery than executing it repeatedly at regular intervals. This observation suggests that it can be advantageous to let dephasing errors accumulate and then efficiently remove them in a single step, rather than attempting frequent incremental corrections. Moreover, we have numerically assessed the steady-state leakage predicted by the perturbative analysis and have confirmed good agreement with the analytical result.

Here, we discuss future directions for our proposal. First, in this work, we have focused on state preparation and error suppression that can be realized using experimentally accessible conditional Gaussian operations. An interesting next step is to extend the autonomous QEC to incorporate additional elements, such as conditional non-Gaussian operations, which may enable the engineering of a broader class of quantum states. Second, while we have studied the preparation of cubic phase and trisqueezed states and the dephasing-error suppression of the (squeezed) cat state, the complementary tasks are also worth exploring: preparing cat and squeezed cat states, and error suppression of the cubic phase and trisqueezed states, as we briefly discuss in Appendix~\ref{app:ProtectionResourceStates}. Finally, it would be interesting to investigate how our nullifier-based digital dissipative synthesis can be combined with other autonomous QEC for squeezed cat qubits that exploit translational symmetry~\cite{shitara2025exploitingtranslationalsymmetryquantum}, enabling not only state stabilization. Such a combination may offer a unified route to engineered dissipation and fault-tolerant control in biased-error bosonic codes, while potentially introducing additional constraints from conditional Gaussian operations and competing timescales among dissipation, control steps, and physical loss.

\section*{Acknowledgments}
This project is supported by Moonshot R\&D, JST, Grant No.\,JPMJMS2061; MEXT Q-LEAP Grant No.\,JPMXS0120319794 and No.\,JPMXS0118068682; JST CREST Grant No.\,JPMJCR23I4; and JST FOREST Grant No.\,JPMJFR223R, Japan.

%\bibliographystyle{apsrev4-1}
%\bibliography{bib}
%merlin.mbs apsrev4-1.bst 2010-07-25 4.21a (PWD, AO, DPC) hacked
%Control: key (0)
%Control: author (72) initials jnrlst
%Control: editor formatted (1) identically to author
%Control: production of article title (-1) disabled
%Control: page (0) single
%Control: year (1) truncated
%Control: production of eprint (0) enabled
%

% -------------------- APPENDICES --------------------
\appendix

\section{Bloch-Messiah decomposition of dissipators}
\label{sppl:Bloch-messiah}

Here, we explain the Bloch-Messiah decompositions to decompose the dissipator into standard Gaussian operations.

\subsection{Cubic phase state}
In this subsection, we prove Eq.~\eqref{eq:Gauss_decomposition}. First, we describe the action of the left-hand-side operator in Eq.~\eqref{eq:Gauss_decomposition} in the Heisenberg picture as
\begin{align}
    \exp&\left[-i\left(\hat{p}-\eta\hat{x}^2\right)t\right]\begin{pmatrix}
            \hat{x}\\ \hat{p}
        \end{pmatrix}\exp\left[i\left(\hat{p}-\eta\hat{x}^2\right)t\right]\nonumber\\
        &=\begin{pmatrix}
            1&0\\-2\eta t&1
        \end{pmatrix}
        \begin{pmatrix}
            \hat{x}\\ \hat{p}
        \end{pmatrix}+
        \begin{pmatrix}
            -t\\\eta t^2
        \end{pmatrix},
\end{align}
with the Baker-Campbell-Hausdorff formula. Then, the coefficient matrix is decomposed into
\begin{align}
    &\begin{pmatrix}
            1&0\\-2\eta t&1
    \end{pmatrix}=
    \begin{pmatrix}
        \cos\phi_\mathrm{CPG,2}(t)&-\sin\phi_\mathrm{CPG,2}(t)\\ \sin\phi_\mathrm{CPG,2}(t)&\cos\phi_\mathrm{CPG,2}(t)
    \end{pmatrix}\nonumber\\
    &\times
    \begin{pmatrix}
        \mathrm{e}^{-r_\mathrm{CPG}(t)}&0\\0&\mathrm{e}^{r_\mathrm{CPG}(t)}
    \end{pmatrix}\nonumber\\
    &\times
    \begin{pmatrix}
        \cos\phi_\mathrm{CPG,1}(t)&-\sin\phi_\mathrm{CPG,1}(t)\\ \sin\phi_\mathrm{CPG,1}(t)&\cos\phi_\mathrm{CPG,1}(t)
    \end{pmatrix},
\end{align}
where $\phi_\mathrm{CPG,1}(t)=\arctan(\sqrt{1+\eta^2t^2}-\eta t)$, $r_\mathrm{CPG}(t)=\ln(\sqrt{1+\eta^2t^2}-\eta t)$, and $\phi_\mathrm{CPG,2}(t)=-\arctan(\sqrt{1+\eta^2t^2}+\eta t)$. Hence, we can implement the operation $\exp\left[i\left(\hat{p}-\eta\hat{x}^2\right)t\right]$ with the rotation, squeezing, another rotation, and displacement operations, as is described in the main text.

\subsection{Trisqueezed state}
In this subsection, we prove Eqs.~\eqref{eq:Gauss_decomposition_TSS}~and~\eqref{eq:Gauss_decomposition_TSS_2}. First, we describe the action of the left-hand-side operator in Eq.~\eqref{eq:Gauss_decomposition_TSS} in the Heisenberg picture as
\begin{align}
    \exp&\left[i\left\{\sqrt{2}\hat{x}-\xi(\hat{x}^2-\hat{p}^2)\right\}t\right]\begin{pmatrix}
            \hat{x}\\ \hat{p}
        \end{pmatrix} \nonumber\\
        &\times\exp\left[-i\left\{\sqrt{2}\hat{x}-\xi(\hat{x}^2-\hat{p}^2)\right\}t\right]\nonumber\\
        &=\begin{pmatrix}
            \cosh(2\xi t)&\sinh(2\xi t)\\\sinh(2\xi t)&\cosh(2\xi t)
        \end{pmatrix}
        \begin{pmatrix}
            \hat{x}\\ \hat{p}
        \end{pmatrix}+\frac{1}{\sqrt{2}\xi}
        \begin{pmatrix}
            1-\cosh(2\xi t)\\-\sinh(2\xi t)
        \end{pmatrix},
\end{align}
with the Baker-Campbell-Hausdorff formula. Then, the coefficient matrix is decomposed into
\begin{align}
    &\begin{pmatrix}
            \cosh(2\xi t)&\sinh(2\xi t)\\\sinh(2\xi t)&\cosh(2\xi t)
    \end{pmatrix}=
    \begin{pmatrix}
        \cos\left(\frac{\pi}{4}\right)&-\sin\left(\frac{\pi}{4}\right)\\ \sin\left(\frac{\pi}{4}\right)&\cos\left(\frac{\pi}{4}\right)
    \end{pmatrix}\nonumber\\
    &\times
    \begin{pmatrix}
        \mathrm{e}^{2\xi t}&0\\0&\mathrm{e}^{-2\xi t}
    \end{pmatrix}\nonumber\\
    &\times
    \begin{pmatrix}
        \cos\left(-\frac{\pi}{4}\right)&-\sin\left(-\frac{\pi}{4}\right)\\ \sin\left(-\frac{\pi}{4}\right)&\cos\left(-\frac{\pi}{4}\right)
    \end{pmatrix}.
\end{align}
Hence, we can implement the operation $\exp\left[i\left\{\sqrt{2}\hat{x}-\xi(\hat{x}^2-\hat{p}^2)\right\}t\right]$ with Gaussian operations.

In the same way, we first describe the action of the left-hand-side operator in Eq.~\eqref{eq:Gauss_decomposition_TSS_2} in the Heisenberg picture as
\begin{align}
    \exp&\left[i\left\{\sqrt{2}\hat{p}+\xi(\hat{x}\hat{p}+\hat{p}\hat{x})\right\}t\right] 
    \begin{pmatrix}
            \hat{x}\\ \hat{p}
        \end{pmatrix}\nonumber\\
        &\times\exp\left[-i\left\{\sqrt{2}\hat{p}+\xi(\hat{x}\hat{p}+\hat{p}\hat{x})\right\}t\right]\nonumber\\
        &=\begin{pmatrix}
        \mathrm{e}^{2\xi t}&0\\0&\mathrm{e}^{-2\xi t}
    \end{pmatrix}
    \begin{pmatrix}
            \hat{x}\\ \hat{p}
        \end{pmatrix}+\frac{1}{\sqrt{2}\xi}
        \begin{pmatrix}
            \mathrm{e}^{2\xi t}-1\\ 0
        \end{pmatrix}.
\end{align}
Hence, we can implement the operation $\exp\left[i\left\{\sqrt{2}\hat{p}+\xi(\hat{x}\hat{p}+\hat{p}\hat{x})\right\}t\right]$ with the squeezing and displacement operators, as is described in the main text.

\subsection{Cat state and squeezed cat state}

In this subsection, we prove Eq.~\eqref{eq:Gauss_decomposition_SqCAT}. First, we describe the action of the left-hand-side operator in Eq.~\eqref{eq:Gauss_decomposition_SqCAT} in the Heisenberg picture as
\begin{align}
    \exp&\left[i\left(\hat{x}^2\mathrm{e}^{2r}-\hat{p}^2\mathrm{e}^{-2r}\right)t\right]\begin{pmatrix}
            \hat{x}\\ \hat{p}
        \end{pmatrix}\exp\left[-i\left(\hat{x}^2\mathrm{e}^{2r}-\hat{p}^2\mathrm{e}^{-2r}\right)t\right]\nonumber\\
        &=\begin{pmatrix}
            \cosh(2t)&-\mathrm{e}^{-2r}\sinh(2t)\\-\mathrm{e}^{2r}\sinh(2t)&\cosh(2t)
        \end{pmatrix}
        \begin{pmatrix}
            \hat{x}\\ \hat{p}
        \end{pmatrix}
\end{align}
with the Baker-Campbell-Hausdorff formula. Then, the coefficient matrix is decomposed into
\begin{align}
    &\begin{pmatrix}
            \cosh(2t)&-\mathrm{e}^{-2r}\sinh(2t)\\-\mathrm{e}^{2r}\sinh(2t)&\cosh(2t)
    \end{pmatrix}\nonumber\\
    &=
    \begin{pmatrix}
        \cos\phi_{\mathrm{SqCAT}}(t)&-\sin\phi_{\mathrm{SqCAT}}(t)\\ \sin\phi_{\mathrm{SqCAT}}(t)&\cos\phi_{\mathrm{SqCAT}}(t)
    \end{pmatrix}\nonumber\\
    &\times
    \begin{pmatrix}
        \mathrm{e}^{-r_{\mathrm{SqCAT}}(t)}&0\\0&\mathrm{e}^{r_{\mathrm{SqCAT}}(t)}
    \end{pmatrix}\nonumber\\
    &\times
    \begin{pmatrix}
        \cos\left(\phi_{\mathrm{SqCAT}}(t)+\frac{\pi}{2}\right)&-\sin\left(\phi_{\mathrm{SqCAT}}(t)+\frac{\pi}{2}\right)\\ \sin\left(\phi_{\mathrm{SqCAT}}(t)+\frac{\pi}{2}\right)&\cos\left(\phi_{\mathrm{SqCAT}}(t)+\frac{\pi}{2}\right)
    \end{pmatrix},
\end{align}
with
\begin{align}
    \tanh(r_{\mathrm{SqCAT}}(t))&=-\frac{\cosh(2r)\sinh(2t)}{\sqrt{1+\cosh^2(2r)\sinh^2(2t)}}\\
    \tan\left(2\phi_{\mathrm{SqCAT}}(t)\right)&=\frac{\cosh(2t)}{\sinh(2r)\sinh(2t)}.
\end{align}
Hence, we can implement the operation $\exp\left[-i\left(\hat{x}^2\mathrm{e}^{2r}-\hat{p}^2\mathrm{e}^{-2r}\right)t\right]$ with the rotation, squeezing, and another rotation operations, as is described in the main text.

\section{Proofs of efficiency for preparing quantum states}\label{app:ProofsOfEfficiency}

In this section, we provide proofs of the theorems stated in Sec.~\ref{Sec:UseCasesAnd}.

We first restate Theorem~1 and provide its proof.

\textbf{Theorem 1 (Preparing a quantum state).}
Let $\ket{\phi}$ be the initial state and $\ket{\psi}$ the target state. Let $\{\ket{\tilde{n}}\}$ be an orthonormal basis with $\ket{\tilde{0}}=\ket{\psi}$ and annihilation operator $\tilde{a}=\hat{\delta}$. Evolve $\ket{\phi}$ under the GKSL dynamics in Eq.~\eqref{eq:Lindblad} for time $\tau$ to obtain $\hat{\rho}(\tau)$. We define a fidelity between quantum states $F(\hat{\rho},\hat{\sigma}):=\mathrm{Tr}\left(\sqrt{\sqrt{\hat{\rho}}\hat{\sigma}\sqrt{\hat{\rho}}}\right)$. To achieve the fidelity $F(\hat{\rho}(\tau),\ket{\psi}\bra{\psi}) = 1 - \epsilon/2$ with $0<\epsilon\ll 1$, the required duration is estimated as
\[
\kappa \tau \simeq \log\!\left(\frac{\langle \tilde{n} \rangle}{\epsilon}\right),
\qquad
\text{where }
\langle \tilde{n} \rangle = \bra{\phi}\tilde{a}^\dagger \tilde{a}\ket{\phi}.
\]
If we approximate the GKSL dynamics via the oscillator--bath map in Eq.~\eqref{eq:LindToDissipation}, the duration satisfies $\kappa\tau \approx N(\Gamma\Delta t)^2$, where $N$ is the circuit depth, $\Gamma$ the coupling strength, and $\Delta t$ the short evolution time. Hence, the depth required to reach fidelity $1-\epsilon/2$ is approximated as
\begin{equation}
    N \simeq \frac{1}{(\Gamma\Delta t)^2}\,
    \log\!\left(\frac{\langle \tilde{n} \rangle}{\epsilon}\right).
\end{equation}

\textit{Proof.}
Let $\hat{\rho}(t)$ denote the state at time $t$ under the GKSL dynamics in Eq.~\eqref{eq:Lindblad}. We define
\begin{equation}
    P_{\tilde{n}}(t) := \bra{\tilde{n}} \hat{\rho}(t) \ket{\tilde{n}},
\end{equation}
which is the population of the level $\ket{\tilde{n}}$ at time $t$. In particular,
\begin{equation}
    P_{\tilde{0}}(\tau)
    = \bra{\tilde{0}} \hat{\rho}(\tau) \ket{\tilde{0}}
    = F\bigl(\hat{\rho}(\tau),\ket{\tilde{0}}\bra{\tilde{0}}\bigr)^2
    = F\bigl(\hat{\rho}(\tau),\ket{\psi}\bra{\psi}\bigr)^2
\end{equation}
is the square of the fidelity of interest.

In the basis $\{\ket{\tilde{n}}\}$, the GKSL master equation in Eq.~\eqref{eq:Lindblad} (with dissipator proportional to $\tilde{a}$) reduces to the rate equation
\begin{equation}\label{eq:diageq}
    \frac{\mathrm{d}P_{\tilde{n}}(t)}{\mathrm{d}t}
    = \kappa\Bigl((\tilde{n}+1)P_{\tilde{n}+1}(t) - \tilde{n}P_{\tilde{n}}(t)\Bigr),
    \qquad \tilde{n}=0,1,2,\dots,
\end{equation}
which expresses the flow of population down the ladder generated by $\tilde{a}$.
This process is known as a special case of the birth-death process, and the solution can be readily obtained~\cite{Norris_1997}.
For completeness, we include the derivation below.

To solve Eq.~\eqref{eq:diageq}, we introduce the generating function
\begin{equation}
    G(x,t) := \sum_{\tilde{n}=0}^{\infty} P_{\tilde{n}}(t)\,x^{\tilde{n}}, \qquad |x|\le 1.
\end{equation}
Using standard identities,
\begin{align}
    \sum_{\tilde{n}=0}^{\infty} (\tilde{n}+1) P_{\tilde{n}+1}(t)\,x^{\tilde{n}}
    &= \frac{\partial G(x,t)}{\partial x},\\
    \sum_{\tilde{n}=0}^{\infty} \tilde{n} P_{\tilde{n}}(t)\,x^{\tilde{n}}
    &= x\,\frac{\partial G(x,t)}{\partial x},
\end{align}
we can rewrite Eq.~\eqref{eq:diageq} as the first-order partial differential equation
\begin{align}
    \frac{\partial G(x,t)}{\partial t}
    &= \sum_{\tilde{n}=0}^{\infty} \kappa\Bigl((\tilde{n}+1)P_{\tilde{n}+1}(t) - \tilde{n}P_{\tilde{n}}(t)\Bigr)x^{\tilde{n}}\nonumber\\
    &= \kappa\left(\frac{\partial G(x,t)}{\partial x} - x\,\frac{\partial G(x,t)}{\partial x}\right)\nonumber\\
    &= \kappa (1-x)\,\frac{\partial G(x,t)}{\partial x}.
\end{align}

We now solve this equation by the method of characteristics. Consider a curve $x = X(t)$ along which $G(X(t),t)$ is constant in time. Along such a curve,
\begin{align}
    \frac{\mathrm{d}}{\mathrm{d}t} G(X(t),t)
    &= \left.\frac{\mathrm{d}X(t)}{\mathrm{d}t}\frac{\partial G(x,t)}{\partial x}\right|_{x=X(t)}
    + \left.\frac{\partial G(x,t)}{\partial t}\right|_{x=X(t)}\nonumber\\
    &= \left[\frac{\mathrm{d}X(t)}{\mathrm{d}t}
        + \kappa\bigl(1 - X(t)\bigr)\right]
       \left.\frac{\partial G(x,t)}{\partial x}\right|_{x=X(t)}.
\end{align}
Requiring $\frac{\mathrm{d}}{\mathrm{d}t} G(X(t),t)=0$ for all $t$ yields the ordinary differential equation
\begin{equation}
    \frac{\mathrm{d}X(t)}{\mathrm{d}t}
    + \kappa\bigl(1 - X(t)\bigr) = 0,
\end{equation}
whose solution is
\begin{align}
    X(t) &= 1 + \bigl(X(0)-1\bigr)\mathrm{e}^{\kappa t}, \\
    \Longleftrightarrow\quad
    X(0) &= 1 + \bigl(X(t)-1\bigr)\mathrm{e}^{-\kappa t}.
\end{align}
Hence, $G$ is transported along characteristics according to
\begin{align}
    G(x,t)
    &= G\bigl(X(t),t\bigr)
     = G\bigl(X(0),0\bigr)\nonumber\\
    &= \sum_{\tilde{n}=0}^{\infty} P_{\tilde{n}}(0)
       \Bigl[1 + (x-1)\mathrm{e}^{-\kappa t}\Bigr]^{\tilde{n}}.
\end{align}

The square of the fidelity of interest is $P_{\tilde{0}}(t)$, which can be obtained from the generating function as
\begin{align}
    P_{\tilde{0}}(t)
    &= G(0,t)\nonumber\\
    &= \sum_{\tilde{n}=0}^{\infty} P_{\tilde{n}}(0)
       \Bigl[1 - \mathrm{e}^{-\kappa t}\Bigr]^{\tilde{n}}.
\end{align}
We now estimate the time $\tau$ required to achieve a fidelity at least $1-\epsilon/2$, i.e., such that
\begin{align}
    \epsilon
    &:= 1 - P_{\tilde{0}}(\tau)
    = 1 - F\bigl(\hat{\rho}(\tau),\ket{\psi}\bra{\psi}\bigr)^2\nonumber\\
    F\bigl(\hat{\rho}(\tau),\ket{\psi}\bra{\psi}\bigr)&\simeq1-\frac{\epsilon}{2}
\end{align}
is small. For small $\epsilon$ we have large $\tau$ and hence $\mathrm{e}^{-\kappa\tau}\ll 1$. In this regime, we expand
\begin{equation}
    1 - \bigl(1 - \mathrm{e}^{-\kappa\tau}\bigr)^{\tilde{n}}
    = \tilde{n} \mathrm{e}^{-\kappa\tau}
      + \mathcal{O}\!\bigl(\mathrm{e}^{-2\kappa\tau}\bigr)
\end{equation}
for each fixed $\tilde{n}$. Using this expansion, the error $\epsilon$ is approximated as
\begin{align}
    \epsilon
    &= 1 - P_{\tilde{0}}(\tau)\nonumber\\
    &= 1 - \sum_{\tilde{n}=0}^{\infty} P_{\tilde{n}}(0)
       \bigl[1 - \mathrm{e}^{-\kappa\tau}\bigr]^{\tilde{n}}\nonumber\\
    &= \sum_{\tilde{n}=0}^{\infty} P_{\tilde{n}}(0)
       \Bigl[1 - \bigl(1 - \mathrm{e}^{-\kappa\tau}\bigr)^{\tilde{n}}\Bigr]\nonumber\\
    &= \sum_{\tilde{n}=0}^{\infty} \tilde{n}\,P_{\tilde{n}}(0)\,\mathrm{e}^{-\kappa\tau}
       + \mathcal{O}\!\bigl(\mathrm{e}^{-2\kappa\tau}\bigr).
\end{align}
The leading term is determined by the mean excitation number of the initial state in the basis $\{\ket{\tilde{n}}\}$,
\begin{equation}
    \sum_{\tilde{n}=0}^{\infty} \tilde{n}\,P_{\tilde{n}}(0)
    = \bra{\phi}\tilde{a}^\dagger\tilde{a}\ket{\phi}
    = \langle\tilde{n}\rangle,
\end{equation}
so that, neglecting terms of order $\mathrm{e}^{-2\kappa\tau}$, we obtain
\begin{align}
    \epsilon
    &\simeq \langle\tilde{n}\rangle\,\mathrm{e}^{-\kappa\tau}\nonumber\\
    \Longleftrightarrow\quad
    \kappa\tau
    &\simeq \log\!\left(\frac{\langle\tilde{n}\rangle}{\epsilon}\right).
\end{align}
This yields the claimed estimate for the time $\tau$ required to achieve fidelity $1-\epsilon/2$ with the target state $\ket{\psi}$.

The statement about the required circuit depth $N$ then follows directly from the approximation $\kappa\tau \approx N(\Gamma\Delta t)^2$ in Eq.~\eqref{eq:LindToDissipation}.
\qed

Hereafter, we discuss the specific application to the cubic phase state and the trisqueezed state, and derive Eqs.~\eqref{eq:ExcitationCPS} and \eqref{eq:ExcitationTSS}. For the cubic phase state, the excitation number $\langle \tilde{n}\rangle_{\mathrm{CPS}}$ can be calculated as
\begin{align}
    \langle \tilde{n}\rangle_{\mathrm{CPS}}&=\bra{0}\hat{\delta}^{\dagger}_{\mathrm{CPS}}(r,\eta)\,\hat{\delta}_{\mathrm{CPS}}(r,\eta)\ket{0}\nonumber\\
    &=\bra{0}\left\{i\frac{\mathrm{e}^{r}}{\sqrt{2}}(\hat{p}-\eta\hat{x}^2)+\frac{\mathrm{e}^{-r}}{\sqrt{2}}\hat{x}\right\}^\dagger\nonumber\\
    &\times\left\{i\frac{\mathrm{e}^{r}}{\sqrt{2}}(\hat{p}-\eta\hat{x}^2)+\frac{\mathrm{e}^{-r}}{\sqrt{2}}\hat{x}\right\}\ket{0}\nonumber\\
    &=\bra{0}\left\{\frac{\mathrm{e}^{2r}}{2}(\hat{p}-\eta\hat{x}^2)^2\right.\nonumber\\
    &\left.+\frac{i}{2}(\hat{x}\hat{p}-\hat{p}\hat{x})+\frac{\mathrm{e}^{-2r}}{2}\hat{x}^2\right\}\ket{0}\nonumber\\
    &=\frac{\mathrm{e}^{2r}}{2}\bra{0}\left(\hat{p}^2+\eta^2\hat{x}^4\right)\ket{0}\nonumber\\
    &+\frac{i}{2}\bra{0}(\hat{x}\hat{p}-\hat{p}\hat{x})\ket{0}+\frac{\mathrm{e}^{-2r}}{2}\bra{0}\hat{x}^2\ket{0}\nonumber\\
    &=\frac{\mathrm{e}^{2r}}{2}\left(\frac{1}{2}+\frac{3}{4}\eta^2\right)+\frac{i}{2}\times i+\frac{\mathrm{e}^{-2r}}{2}\frac{1}{2}\nonumber\\
    &=\frac{1}{2}\cosh(2r)+\frac{3}{8}\eta^{2}\mathrm{e}^{2r}-\frac{1}{2}.
\end{align}
In the same manner, the excitation number of the trisqueezed state in Eq.~\eqref{eq:ExcitationTSS} can be calculated as
\begin{align}
    \langle \tilde{n}\rangle_{\mathrm{TSS}}&=\bra{0}\hat{\delta}^{\dagger}_{\mathrm{TSS}}(\xi)\,\hat{\delta}_{\mathrm{TSS}}(\xi)\ket{0}\nonumber\\
    &=\bra{0}\left\{\hat{a}-\xi(\hat{a}^\dagger)^2+\mathcal{O}(\xi^2)\right\}^\dagger\nonumber\\
    &\times\left\{\hat{a}-\xi(\hat{a}^\dagger)^2+\mathcal{O}(\xi^2)\right\}\ket{0}\nonumber\\
    &=2\xi^2+\mathcal{O}(\xi^3).
\end{align}
Hence, we can derive Eqs.~\eqref{eq:ExcitationCPS} and \eqref{eq:ExcitationTSS}.

\section{Numerical verification of Trotter approximation accuracy}\label{app:NumericalVerificationOf}
\begin{figure}[t]
    \centering
    \includegraphics[width=0.9\linewidth]{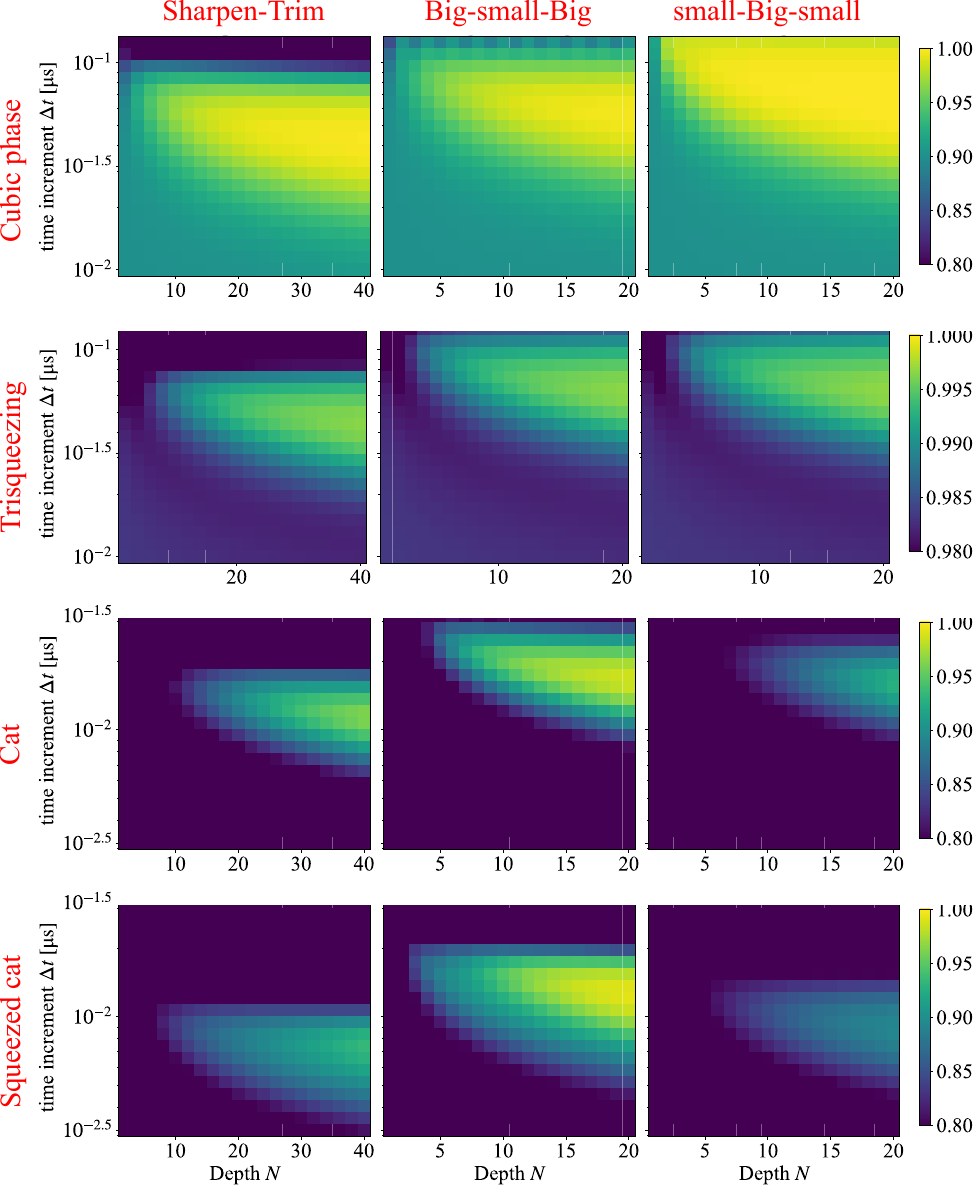}
    \caption{\textbf{Autonomous QEC performance as a function of the circuit depth $N$ and the time increments $\Delta t$.} Each row corresponds to a different quantum state, namely the cubic phase state, trisqueezed state, cat state, and squeezed cat state, while each column represents a different QEC protocol, namely the Sharpen--Trim, Big--small--Big, and small--Big--small schemes. In each panel, the horizontal axis is the depth $N$, the vertical axis is the time step $\Delta t$, and the color scale indicates the fidelity.
    }
    \label{fig:QEC_grid}
\end{figure}

In this section, we numerically assess the accuracy of our autonomous QEC as a function of the circuit depth $N$ and the time increments $\Delta t$. In Fig.~\ref{fig:QEC_grid}, we summarize the results for all combinations of four target states and three QEC protocols. Each row corresponds to a different quantum state, namely the cubic phase state, trisqueezed state, cat state, and squeezed cat state, while each column represents a different QEC protocol, namely the Sharpen--Trim, Big--small--Big, and small--Big--small schemes. In each panel, the horizontal axis is the depth $N$, the vertical axis is the time step $\Delta t$, and the color scale indicates the fidelity. For the preparation tasks (cubic phase and trisqueezed states), we prepare the vacuum state, apply a single round of the autonomous QEC with parameters $(N,\Delta t)$, and evaluate the fidelity with the ideal target state. For the error-suppression tasks (cat and squeezed cat states), we instead start from the ideal target state, evolve it under the dephasing error channel for a fixed duration, apply the same QEC protocol, and compute the fidelity with respect to the initial state.

The simulation parameters in Fig.~\ref{fig:QEC_grid} are aligned with those used in Fig.~\ref{fig:QEC_generating}. As target states, we consider a cubic phase state with squeezing level $5~\mathrm{dB}$ and cubicity $\eta=0.3$, a trisqueezed state with trisqueezing level $2~\mathrm{dB}$, a cat state with amplitude $\alpha=3$, and a squeezed cat state with amplitude $\alpha=3$ and squeezing level $5~\mathrm{dB}$. For the cat and squeezed cat states, we consider the dephasing error with $\hat{L}=\hat{a}^\dagger\hat{a}$. The error rate is fixed at $\kappa=5~\mathrm{kHz}$, and the system undergoes noisy evolution for $200~\mathrm{\mu s}$ prior to the QEC round. During the QEC operations, we also include ancillary-qubit decoherence with $T_1=T_2=100~\mathrm{\mu s}$, and we assume a QEC gate speed $\Gamma$ of $10~\mathrm{MHz}$.

As expected, smaller $\Delta t$ and larger $N$ generally improve the performance, reflecting the convergence of the Trotterized implementation of the QEC dynamics. Conversely, for overly large $\Delta t$, Trotterization errors accumulate and reduce the fidelity in all protocols. Based on the observations in Fig.~\ref{fig:QEC_grid}, we adopt the small--Big--small protocol for the cubic phase state and the trisqueezed state, and the Big--small--Big protocol for the cat state and the squeezed cat state. We also adopt the corresponding $(N, \Delta t)$ chosen from the high-fidelity region of Fig.~\ref{fig:QEC_grid} for the time-series comparisons presented in Fig.~\ref{fig:QEC_protecting}.

\section{Error suppression of resource states}
\label{app:ProtectionResourceStates}

\begin{figure}[t]
    \centering
    \includegraphics[width=1.0\linewidth]{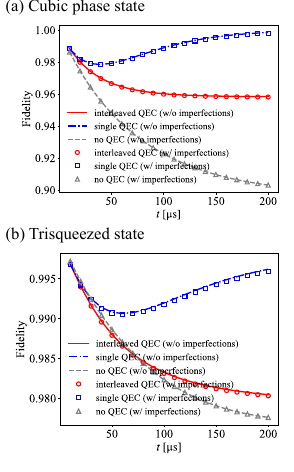}
    \caption{\textbf{Numerical simulations of suppressing errors of resource states.}
    Time evolution of the fidelity between the ideal target state and the noisy state under three strategies:
    no error correction (``no QEC''), a single application of the autonomous QEC immediately before readout (``single QEC''), and repeated applications during storage (``interleaved QEC''). Here, we consider the photon loss error. Lines show the idealized numerical results without imperfections during the QEC operations, whereas points show the results including such imperfections. (a) Cubic phase state with squeezing level $5\ \mathrm{dB}$ and $\eta=0.3$. (b) Trisqueezed state with trisqueezing level $2\ \mathrm{dB}$.
    }
    \label{fig:QEC_protecting_resource}
\end{figure}

We complement the error-suppression simulations in Sec.~\ref{sec:numerics}, which focused on the (squeezed) cat state, by repeating the similar analysis for the non-Gaussian resource states studied in Sec.~\ref{sec:cv_states}: the cubic phase state and the trisqueezed state. The simulation protocol follows Fig.~\ref{fig:QEC_protecting}. The horizontal axis represents the storage time under a fixed error process, and the vertical axis represents the fidelity between the ideal target state and the corresponding simulated state. For our autonomous QEC, we compare three strategies: (i) passive storage without QEC (``no QEC''), (ii) a single execution of our autonomous QEC immediately before readout (``single QEC''), and (iii) repeated shallow executions interleaved with noisy storage (``interleaved QEC''). The results are summarized in Fig.~\ref{fig:QEC_protecting_resource}.

We now describe the simulation model and parameters used in Fig.~\ref{fig:QEC_protecting_resource}. We model the storage error by the GKSL equation in Eq.~\eqref{eq:GKSL} with the photon loss error $\hat{L}=\hat{a}$ and rate $\kappa=5~\mathrm{kHz}$. The target states are (a) a cubic phase state with cubicity $\eta=0.3$ and squeezing level $5~\mathrm{dB}$, and (b) a trisqueezed state with trisqueezing level $2~\mathrm{dB}$. For the digitized implementation of the autonomous QEC, we use the small--Big--small protocol with gate rate $\Gamma=10~\mathrm{MHz}$, and we include ancillary-qubit decoherence with $T_1=T_2=100~\mathrm{\mu s}$ during the QEC operations. In the ``interleaved QEC'' strategy, we apply 20 rounds uniformly over $200~\mathrm{\mu s}$, with $(N,\Delta t)=(1,0.05~\mathrm{\mu s})$ per round. In contrast, for the ``single QEC'' protocol, we apply the QEC protocol only once, immediately before the readout, with the same time increment $\Delta t = 0.01~\mathrm{\mu s}$. At each readout time, the depth $N$ is set equal to the number of QEC rounds that would have been applied in the ``interleaved QEC'' protocol up to that time, so that the total QEC depth is the same in the two strategies.

As shown in Fig.~\ref{fig:QEC_protecting_resource}, the autonomous QEC suppresses the error of both resource states under photon loss: the fidelities achieved by ``single QEC'' and ``interleaved QEC'' stay above those of ``no QEC'' for sufficiently large $t$. For the parameters considered here, applying the correction once immediately before readout yields the largest recovery, consistent with the trend observed for the error suppression of the (squeezed) cat state in Sec.~\ref{sec:numerics}.

\section{Perturbative analysis of dephasing-induced leakage in the squeezed cat code space}
\label{app:perturb-sqcat}

In this section, we derive a first-order (in error rate) expression for the leakage from the squeezed cat code space under the proposed autonomous QEC and dephasing error. The calculation reduces to solving a constrained linear system in Liouville space, which is well-posed once parity conservation is used to restrict the dynamics to the even-parity sector.

\subsection{Settings}
We consider a single bosonic mode under dynamics governed by the GKSL master equation with vanishing Hamiltonian ($\hat{H}=0$):
\begin{align}
\od{1}{\hat{\rho}}{t}=\kappa_{\rm s}\,\mathcal{D}[\hat{\delta}_{\mathrm{SqCAT}}]\hat{\rho}+\kappa_\phi\,\mathcal{D}[\hat{a}^\dagger \hat{a}]\hat{\rho},
\label{eq:app_master_equation}
\end{align}
where $\kappa_{\rm s}$ is the dissipative stabilization rate, and $\kappa_\phi$ is the dephasing rate. We omit the arguments $(\alpha,r)$ of the nullifier $\hat{\delta}_{\mathrm{SqCAT}}$ throughout this section. When $\kappa_\phi=0$, the dynamics drives the state into the squeezed cat code space. We define the perturbation parameter
\begin{align}
\epsilon \coloneq \frac{\kappa_\phi}{\kappa_{\rm s}}\ll 1,
\label{eq:app_eps_def}
\end{align}
and write the Liouvillian as $\mathcal{L}=\kappa_{\rm s}(\mathcal{L}_0+\epsilon\mathcal{L}_1)$ with
\begin{align}
\mathcal{L}_0 \coloneq \mathcal{D}[\hat{\delta}_{\mathrm{SqCAT}}],
\quad
\mathcal{L}_1 \coloneq \mathcal{D}[\hat{a}^\dagger\hat{a}].
\label{eq:app_L0_L1_def}
\end{align}

We define the projector onto the two-dimensional code space by
\begin{align}
\hat{P}_{\rm code} \coloneq \ket{\mathrm{SqCAT}_{+}}\bra{\mathrm{SqCAT}_{+}} + \ket{\mathrm{SqCAT}_{-}}\bra{\mathrm{SqCAT}_{-}}.
\label{eq:app_Pcode_def}
\end{align}
In the main text, we quantify leakage as the weight outside the code subspace,
\begin{align}
w(\hat{\rho})\coloneq \mathrm{Tr}\!\left[(\hat{I}-\hat{P}_{\rm code})\hat{\rho}\right].
\label{eq:app_leakage_def}
\end{align}
In particular, we consider the leakage of the steady state $\hat{\rho}_{\rm ss}$ when we start from the even squeezed cat initial state
\begin{align}
\hat{\rho}(0)=\ket{\mathrm{SqCAT}_{+}}\bra{\mathrm{SqCAT}_{+}}.
\label{eq:app_even_initial}
\end{align}

\subsection{Parity restriction}
Before performing a perturbative analysis of the steady state, we clarify the parity structure in the dynamics.
We define the photon-number parity operator $\hat{\Pi}\coloneq  {\rm e}^{i\pi \hat{a}^\dagger\hat{a}}=\hat{P}_{\rm e}-\hat{P}_{\rm o}$, where $\hat{P}_{\rm e}$ and $\hat{P}_{\rm o}$ are the projections onto the even and odd parity subspace defined by
\begin{align}
    \hat{P}_{\rm e}&=\sum_{n=0}^\infty \ket{2n}\bra{2n},\\
    \hat{P}_{\rm o}&=\sum_{n=0}^\infty \ket{2n+1}\bra{2n+1},
\end{align}
respectively.
The squeezed cat states are the eigenstates of the parity operator: $\hat{\Pi}\ket{\mathrm{SqCAT}_{\pm}}=\pm \ket{\mathrm{SqCAT}_{\pm}}$.
We also note that both the dephasing error operator $\hat{a}^\dagger\hat{a}$ and the nullifier $\hat{\delta}_{\mathrm{SqCAT}}$ commute with the parity operator.
Therefore, the dynamics described by the master equation~\eqref{eq:app_master_equation} preserve the parity.
More formally, we define a superoperator ${\mathcal P}_{i,j}(\cdot)=\hat{P}_i\cdot\hat{P}_j$ for $i,j={\rm e,o}$.
Then, the full Liouvillian is block diagonal in each parity sector:
\begin{align}
    \mathcal{L}=\bigoplus_{i,j={\rm e,o}} {\mathcal P}_{i,j}\circ \mathcal{L}\circ {\mathcal P}_{i,j}.
    \label{eq:app_L_blockdiag}
\end{align}
Since we start from the even squeezed cat initial state satisfying $\hat{\rho}(t)={\mathcal P}_{\rm e,e}(\hat{\rho}(0))$, the state remains in the even sector for all times, $\hat{\rho}(t)={\mathcal P}_{\rm e,e}(\hat{\rho}(t))$ and $\hat{\rho}_{\rm ss}={\mathcal P}_{\rm e,e}(\hat{\rho}_{\rm ss})$.
Therefore, both the transient dynamics and the steady-state problem can be restricted to $\mathcal{L}^{(\mathrm{even})}\coloneq{\mathcal P}_{\rm e,e}\circ \mathcal{L} \circ{\mathcal P}_{\rm e,e}$.
We will use this fact later in this section.

\subsection{Perturbative expansion of the steady state}
The steady state of the master equation~\eqref{eq:app_master_equation} satisfies 
\begin{align}
    {\mathcal L_0(\hat{\rho}_{\rm ss}) + \epsilon \mathcal L_1}(\hat{\rho}_{\rm ss})=0.
    \label{eq:app_steady}
\end{align}
To determine the steady state $\hat{\rho}_{\rm ss}$ perturbatively, we expand it in powers of $\epsilon$ as
\begin{align}
\hat{\rho}_{\rm ss}=\hat{\rho}_0+\epsilon \hat{\rho}_1+O(\epsilon^2).
\label{eq:app_rho_expansion}
\end{align}
Then, we substitute Eq.~\eqref{eq:app_rho_expansion} to Eq.~\eqref{eq:app_steady} and solve it order by order.
At zeroth order, we obtain $\mathcal{L}_0\hat{\rho}_0=0$, whose solution can be any density operator supported on the code space. %, ${\rm span}\{\ket{\mathrm{SqCAT}_{+}},\ket{\mathrm{SqCAT}_{-}}\}$. 
Since the initial state~\eqref{eq:app_even_initial} is in the even parity sector and the dynamics preserves the parity, the steady state is also in the even parity sector:
\begin{align}
\hat{\rho}_0=\ket{\mathrm{SqCAT}_{+}}\bra{\mathrm{SqCAT}_{+}}.
\label{eq:app_rho0_def}
\end{align}
At first order, we obtain
\begin{align}
\mathcal{L}_0\hat{\rho}_1=-\mathcal{L}_1\hat{\rho}_0.
\label{eq:app_first_order_equation}
\end{align}
The superoperator $\mathcal{L}_0$ has a 4-dimensional kernel as
\begin{align}
    \mathcal{L}_0(\ket{\mathrm{SqCAT}_{k}}\bra{\mathrm{SqCAT}_{l}})=0 \quad (k,l=\pm)
\end{align}
and hence is non-invertible, so some additional condition is required to solve Eq.~\eqref{eq:app_first_order_equation}.
If the dimension of the kernel of $\mathcal{L}_0$ is one, the trace condition $\Tr [\hat{\rho}_i]=0 \ (i\ge1)$, which originates from the trace-preserving property of the dynamics, suffices to uniquely determine $\hat{\rho}_1$. To make this condition sufficient in the present setting, we restrict the whole Hilbert space to the even-parity subspace, where the dimension of the kernel is reduced from four to one. Namely, $\mathcal{L}_0^{(\mathrm{even})} \coloneq {\mathcal P}_{\rm e,e}\circ \mathcal{L}_0 \circ{\mathcal P}_{\rm e,e}$ has a one-dimensional kernel spanned by $\ket{\mathrm{SqCAT}_{+}}\bra{\mathrm{SqCAT}_{+}}$, and  $\hat{\rho}_1$ can be uniquely determined by the following equations:
\begin{align}
    \mathcal{L}_0^{(\mathrm{even})}\hat{\rho}_1&=-\mathcal{L}_1^{(\mathrm{even})}\hat{\rho}_0,   \label{eq:app_superoprator_rep1}\\
    \Tr[\hat{\rho}_1]&=0. \label{eq:app_superoprator_rep2}
\end{align}

In numerical calculation, we vectorize operators as $\hat{\rho}\rightarrow\ket{\rho}\!\rangle$, where the inner product coincides with the Hilbert-Schmidt inner product in the original operator space as $\langle\!\braket{A}{B}\!\rangle=\Tr[\hat{A}^\dagger\hat{B}]$.
Accordingly, superoperators are expressed as an operator acting on the vectorized operator, $\mathcal{L}_i^{(\mathrm{even})}\rightarrow \mathbf{L}_i^{(\mathrm{even})} \ (i=0,1)$.
Noting that the trace condition can be expressed as $0=\Tr[\hat{\rho}_1]=\Tr[\hat{I}\hat{\rho}_1]=\langle\!\braket{I}{{\rho}_1}\!\rangle$, Eqs.~\eqref{eq:app_superoprator_rep1} and~\eqref{eq:app_superoprator_rep2} can be rewritten as
\begin{align}
\begin{pmatrix}
\mathbf{L}_0^{(\mathrm{even})}\\
\langle\!\langle I|
\end{pmatrix}
\ket{\rho_1}\!\rangle=
\begin{pmatrix}
-\mathbf{L}_1^{(\mathrm{even})}\ket{\rho_0}\!\rangle\\
0
\end{pmatrix}
\label{eq:app_augmented_system}
\end{align}
We numerically solve this augmented linear system in the Fock basis.

\subsection{Leakage coefficient $A(\alpha,r)$}
\begin{figure}[htbp]
    \centering
    \includegraphics[width=0.9\linewidth]{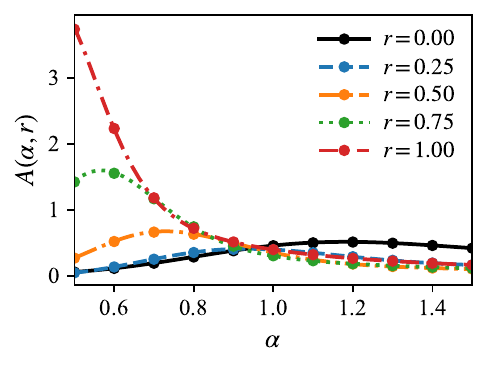}
    \caption{First-order dephasing-induced leakage coefficient $A(\alpha,r)$ for the dissipatively stabilized even squeezed cat code space as a function of the cat amplitude $\alpha$, shown for several squeezing parameters $r$.}
    \label{fig:QEC_leakage_coefficient}
\end{figure}
We have obtained the perturbative expansion of the steady state up to the first order.
By substituting it into the definition of the steady-state leakage, we obtain
\begin{align}
w(\hat{\rho}_{\rm ss})= \epsilon A(\alpha,r)+O(\epsilon^2),
\label{eq:app_wout_expansion}
\end{align}
where the first-order coefficient can be expressed as $A(\alpha,r):= -\mathrm{Tr}\!\left(\hat{P}_{\rm code}\hat{\rho}_1\right)$.
In Fig.~\ref{fig:QEC_leakage_coefficient}, we plot the leakage coefficient $A(\alpha, r)$ against $\alpha$ for several values of the squeezing parameter $r$.
The lines represent the perturbative evaluation of $A(\alpha,r)$ using the method described above.
The points represent the estimates of 
$A(\alpha,r)$ obtained by computing the steady state through long-time evolution according to the master equation~\eqref{eq:app_master_equation} for various values of $\epsilon$, and then extracting the coefficient from a linear regression of $w(\hat{\rho}_{\rm ss})$ against $\epsilon$. We see that the two estimates obtained by different methods are in good agreement.
For each fixed squeezing parameter $r$, $A(\alpha,r)$ is non-monotonic in $\alpha$ and exhibits a peak at an intermediate cat amplitude $\alpha$. As $r$ is increased, the location of this maximum shifts to a smaller $\alpha$.

\end{document}